\documentclass[a4paper,11pt]{article}

\usepackage{jheppub} 

\usepackage[T1]{fontenc} 

\author[a,b]{Marco Bochicchio}
\affiliation[a]{INFN sez. Roma 1\\Piazzale A. Moro 2, Roma, I-00185, Italy}
\affiliation[b]{Scuola Normale Superiore (SNS)\\Piazza dei Cavalieri 7, Pisa, I-56100, Italy}

\emailAdd{marco.bochicchio@roma1.infn.it}

\abstract{We find an asymptotic solution for two- and three-point correlators of local gauge-invariant operators, in a lower-spin sector of massless large-$N$ $QCD$, in terms of glueball and meson propagators, by means of a new purely field-theoretical technique that we call the asymptotically-free bootstrap. 
The asymptotically-free bootstrap exploits the lowest-order conformal invariance of connected correlators of gauge invariant composite operators in perturbation theory, the renormalization-group improvement, and a recently-proved asymptotic structure theorem for glueball and meson propagators, that involves the unknown particle spectrum and the anomalous dimension of operators for fixed spin. In principle the asymptotically-free bootstrap extends to all the higher-spin two- and three-point correlators whose lowest-order conformal limit is non-vanishing in perturbation theory, and by means of the operator product expansion to the corresponding asymptotic multi-point correlators as well. Besides, the asymptotically-free bootstrap provides asymptotic three-point (and to some extent also multi-point) $S$-matrix amplitudes in massless large-$N$ $QCD$ in terms of glueball and meson propagators as opposed to perturbation theory. Remarkably, the asymptotic $S$-matrix depends only on the unknown particle spectrum, but not on the anomalous dimensions, as a consequence of the $LSZ$ reduction formulae. Moreover, the asymptotically-free bootstrap applies to large-$N$ $\mathcal{N}=1$ $SUSY$ $YM$ as well. Very many physics consequences follow, both practically and theoretically. Practically, as just a few examples among many more, it follows the structure of the light by light scattering amplitude, of the pion form factor, and the associated vector dominance. Theoretically, the asymptotic solution sets the strongest constraints on any actual solution of large-$N$ $QCD$ (and of large-$N$ $\mathcal{N}=1$ $SUSY$ $YM$), and in particular on any string solution.}

\usepackage{braket}
\usepackage{amsmath}
\usepackage{amssymb}
\usepackage{graphicx}
\usepackage{hyperref}
\usepackage{fancyhdr}
\newcommand{\Lambdams}{\Lambda_{\overline{MS}}}

\newcommand{\plms}{\frac{p^2}{\Lambdams^2}}

\def\beq{\begin{equation}}
\def\eeq{\end{equation}}
\def\bea{\begin{eqnarray}}
\def\eea{\end{eqnarray}}
\def\bq{\begin{quote}}
\def\eq{\end{quote}}
\DeclareMathOperator{\Tr}{Tr}

\title{An asymptotic solution of large-$N$ $QCD$, \\
and of large-$N$ $\mathcal{N}=1$ $SUSY$ $YM$  \\}
\date{}
\begin{document}
\maketitle
%
%
%


\section{Introduction and results} \label{s0}

Solving $QCD$ in 't Hooft large-$N$ limit \cite{H1} is a long-standing difficult problem. An easier problem is to find a solution, not exact, but only asymptotic in the ultraviolet ($UV$). \par
In a sense this asymptotic solution in the $UV$ already exists: It is ordinary perturbation theory. \par
But in fact it is much more interesting an asymptotic solution in the $UV$ written in terms of glueballs and mesons as opposed to gluons and quarks. \par
An asymptotic solution of this kind would replace $QCD$ viewed as a theory of gluons and quarks, that are strongly coupled in the infrared in perturbation theory, with a theory of an infinite number of glueballs and mesons, that are weakly coupled at all scales in the large-$N$ limit.\par
As it is well known, the last statement follows by standard large-$N$ estimates for Feynman graphs \cite{H1} in 't Hooft limit of large-$N$ $QCD$, i.e. in the large-$N$ limit of $SU(N)$ Yang-Mills ($YM$) with $N_f$ quarks in the fundamental representation, with $N_f=const$ as $N$ diverges. \par
For example, in the pure-glue sector connected $r$-point correlators of local single-trace gauge-invariant operators scale as $N^{2-r}$. Therefore, at leading $\frac{1}{N}$ order only one-point condensates do not vanish, while at next-to-leading order only two-point correlators occur. But since at this order the interaction associated to three- and multi-point correlators vanishes, by the Kallen-Lehmann representation the two-point correlators must be a sum of propagators of free fields, involving single-particle pure poles \cite{Mig}. At the next order the interaction arises, but it is parametrically weak in the $\frac{1}{N}$ expansion.  \par
On the contrary, mesons and glueballs in Veneziano limit \cite{Veneziano0} of $QCD$, defined by $\frac{N_f}{N}=const$ as $N$ diverges, have widths on the order of a power of $\frac{N_f}{N}$, i.e. on the order of $1$ (see \cite{Veneziano0} for detailed estimates both in 't Hooft and Veneziano limit), and therefore the two-point correlators admit a Kallen-Lehmann representation that involves a multi-particle continuum \cite{Veneziano0} and not just single-particle pure poles. As a consequence neither the asymptotic theorem described in the following, nor the main results of this paper apply to Veneziano limit in the form reported in this paper. \par
Recently, the asymptotic structure of two-point correlators for any integer spin has been explicitly characterized by the asymptotic theorem \cite{MBN} reported below, in 't Hooft limit of any large-$N$ confining asymptotically-free gauge theory massless in perturbation theory satisfying the properties listed in the following, such as
massless $QCD$ (i.e. $QCD$ with massless quarks) and $\mathcal{N}=1$ supersymmetric ($SUSY$) $YM$. \par
To avoid ambiguities in this more general framework, we make more explicit our basic assumptions, that capture the fundamental features of 't Hooft limit of large-$N$ massless $QCD$. \par
Firstly, we assume that our large-$N$ gauge theory confines, by which we mean that the one-particle mass spectrum for fixed spin is a discrete diverging sequence at the leading large-$N$ order. In fact, this assumption is not independent on the following one. \par
Secondly, we assume that our gauge theory has a large-$N$ limit of 't Hooft type, i.e. that the Kallen-Lehmann representation of the connected two-point correlators that we are interested in is saturated by a sum of free propagators, because by assumption the interaction and the particle widths are suppressed at the leading large-$N$ order. \par 
Thirdly, we assume that our gauge theory is asymptotically-free and massless to all orders in perturbation theory, and that it admits a conformal limit at lowest order in perturbation theory. \par 
Under these assumptions the asymptotic theorem described below holds. \par
The asymptotic theorem for the two-point correlators is the basis of a new technique described in this paper, that we call the asymptotically-free bootstrap \footnote{The name derives by the celebrated conformal bootstrap.}, by which we extend the asymptotic theorem to three-point and, to some extent, to multi-point correlators
and $S$-matrix amplitudes, getting in this way an asymptotic solution of large-$N$ $QCD$ (and of $\mathcal{N}$ $=1 $ $SUSY$ $YM$) in a sense specified below. The basic ideas underlying the asymptotically-free bootstrap have been already described in \cite{QCD0}, in this paper we furnish detailed arguments. \par
The asymptotic theorem for two-point correlators is based on the Callan-Symanzik equation, plus the Kallen-Lehmann representation, plus the assumption that the theory confines, i.e. technically that the one-particle spectrum for integer spin $s$ at the leading $\frac{1}{N}$ order is a discrete diverging sequence with smooth asymptotic distribution $ \rho_s(m^2)$. Its proof, according to \cite{MBN}, is reported in section \ref{s11}, since in this paper we need some refinements of it and its further extension to the coefficients of the operator product expansion ($OPE$). \par
In this introduction we recall the precise statement of the asymptotic theorem, because it is necessary to explain the logic of this paper: \\
The connected two-point Euclidean correlator of a local gauge-invariant single-trace operator (and of a fermion bilinear) $\mathcal{O}^{(s)}$ of integer spin $s$ and naive mass dimension $D$ and with anomalous dimension $\gamma_{\mathcal{O}^{(s)}}(g)$,
must factorize asymptotically for large momentum, and at the leading order in the large-$N$ limit, over the following poles and residues (after analytic continuation to Minkowski space-time):
\bea \label{at1}
\int \langle \mathcal{O}^{(s)}(x) \mathcal{O}^{(s)}(0) \rangle_{conn}\,e^{-ip\cdot x}d^4x
\sim \sum_{n=1}^{\infty}  P^{(s)} \big(\frac{p_{A}}{m^{(s)}_n}\big) \frac{m^{(s)2D-4}_n Z_n^{(s)2}  \rho_s^{-1}(m^{(s)2}_n)}{p^2+m^{(s)2}_n  } \nonumber \\
\eea
where $ P^{(s)} \big( \frac{p_{A}}{m^{(s)}_n} \big)$ is a dimensionless polynomial in the four momentum $p_{A}$ \footnote{We employ latin letters $A,\cdots$ to denote vector indices, and greek letters $\alpha, \dot \alpha, \cdots$ to denote spinor indices.} that projects on the free propagator of spin $s$ and mass $m^{(s)}_n$ and:
\bea \label{g}
\gamma_{\mathcal{O}^{(s)}}(g)= - \frac{\partial \log Z^{(s)}}{\partial \log \mu}=-\gamma_{0} g^2 + O(g^4)
\eea 
with $Z_n^{(s)}$ the associated renormalization factor computed on shell, i.e. for $p^2=m^{(s)2}_n$:
\bea \label{z}
Z_n^{(s)}\equiv Z^{(s)}(m^{ (s)}_n)= \exp{\int_{g (\mu)}^{g (m^{(s)}_n )} \frac{\gamma_{\mathcal{O}^{(s)}} (g)} {\beta(g)}dg}
\eea
The symbol $\sim$ means always in this paper asymptotic equality in the sense specified below, up to perhaps a constant factor overall. \par
The sum in the right-hand side ($RHS$) of Eq.(\ref{at1}) is in fact badly divergent, but the divergence is a contact term, i.e. a polynomial of finite degree in momentum. Thus the infinite sum in the $RHS$ of Eq.(\ref{at1}) makes sense only after subtracting the contact terms (see remark below Eq.(\ref{at2})). Fourier transforming Eq.(\ref{at1}) in the coordinate representation
projects away for $x\neq 0$ the contact terms and avoids convergence problems:
\bea \label{at0}
\langle \mathcal{O}^{(s)}(x) \mathcal{O}^{(s)}(0) \rangle_{conn} 
\sim \sum_{n=1}^{\infty} \frac{1}{(2 \pi)^4} \int  P^{(s)} \big(\frac{p_{A}}{m^{(s)}_n}\big) \frac{m^{(s)2D-4}_n Z_n^{(s)2} \rho_s^{-1}(m^{(s)2}_n)}{p^2+m^{(s)2}_n  } \,e^{ip\cdot x}d^4p \nonumber \\
\eea
The proof of the asymptotic theorem reduces to showing that Eq.(\ref{at1})
matches asymptotically for large momentum, within the universal leading and next-to-leading logarithmic accuracy,
the renormalization-group ($RG$) improved perturbative result implied \footnote{ This asymptotic estimate holds for every $\gamma'$ but $0,1$ (see section \ref{s11}).} by the Callan-Symanzik equation:
\bea \label{CS}
&& \int \langle \mathcal{O}^{(s)}(x) \mathcal{O}^{(s)}(0) \rangle_{conn}\,e^{-ip\cdot x}d^4x  \nonumber \\
&& \sim P^{(s)}\big(\frac{p_{A}}{p}\big) \, p^{2D-4}    \Biggl[\frac{1}{\beta_0\log (\frac{p^2}{\Lambda^2_{QCD} } )}\biggl(1-\frac{\beta_1}{\beta_0^2}\frac{\log\log (\frac{p^2}{\Lambda^2_{QCD} } ) }{\log (\frac{p^2}{\Lambda^2_{QCD} } )}    + O(\frac{1}{\log (\frac{p^2}{\Lambda^2_{QCD} } )} ) \biggr)\Biggr]^{\gamma'-1}
\eea
up to contact terms (i.e. distributions supported at coinciding points), and that this matching fixes uniquely the universal asymptotic behavior of the residues in Eq.(\ref{at1}), with
$ \gamma'=\frac{\gamma_0}{\beta_0}$, $\beta_0$, $\beta_1$, $\gamma_0$ the first and second coefficients of the beta function and the first coefficient of the anomalous dimension respectively, and $\Lambda_{QCD}$
the $QCD$ \footnote{We refer mainly to $QCD$ with massless quarks, but in fact the asymptotically-free bootstrap applies to $\mathcal{N}$ $=1$ $SUSY$ $YM$, and more generally to any large-$N$ confining asymptotically-free gauge theory massless in perturbation theory satisfying the aforementioned assumptions.} $RG$-invariant scale in some scheme. 
More precisely, the asymptotic behavior of the residues is fixed by the asymptotic theorem within the universal, i.e. the scheme-independent, leading and next-to-leading logarithmic accuracy.
This implies that the renormalization factors in the residues are fixed asymptotically for large $n$ to be:
\begin{equation}\label{z1}
Z_n^{(s)2}\sim 
\Biggl[\frac{1}{\beta_0\log \frac{ m^{ (s) 2}_n }{ \Lambda^2_{QCD} }} \biggl(1-\frac{\beta_1}{\beta_0^2}\frac{\log\log \frac{ m^{ (s) 2}_n }{ \Lambda^2_{QCD} }}{\log \frac{ m^{ (s) 2}_n }{ \Lambda^2_{QCD} }}    + O(\frac{1}{\log \frac{ m^{ (s) 2}_n }{ \Lambda^2_{QCD} } } ) \biggr)\Biggr]^{\gamma'}
\end{equation}
Eq.(\ref{at1}) for the correlator can be rewritten equivalently as:
\bea \label{at2}
\int \langle \mathcal{O}^{(s)}(x) \mathcal{O}^{(s)}(0) \rangle_{conn}\,e^{-ip\cdot x}d^4x  
\sim    P^{(s)} \big(\frac{p_{A}}{p} \big)  \, p^{2D-4}   \sum_{n=1}^{\infty} \frac{Z_n^{(s)2}   \rho_s^{-1}(m^{(s)2}_n)  }{p^2+m^{(s)2}_n  }
\eea
up to (divergent) contact terms, where now the sum in the $RHS$ is convergent for $\gamma'> 1$. Otherwise, it is divergent but the divergence is again a contact term. $P^{(s)} \big(\frac{p_{A}}{p} \big)$ is the projector obtained substituting $-p^2$  to $m_n^2$ in  $P^{(s)} \big(\frac{p_{A}}{m_n} \big)$\footnote{We use Veltman conventions for Euclidean and Minkowskian propagators of spin $s$ \cite{Velt2}.}. \par
An important corollary \cite{MBN} of the asymptotic theorem is that to compute the asymptotic behavior we need not to know explicitly neither the actual spectrum nor the asymptotic spectral distribution, since it cancels by evaluating the sum in Eq.(\ref{at2}) by the integral that occurs as the leading term in Euler-MacLaurin formula (see section \ref{s11}). Hence $RG$-improved perturbation theory does not contain in fact spectral information \cite{MBN}, as perhaps expected, and so it does not our asymptotic solution. \par
In order to get spectral information it is necessary to lift the asymptotic solution to the actual solution. 
We discuss ideas in this direction in section \ref{s1}.  \par
Nevertheless, from a practical point of view, the asymptotic formulae suitably interpreted can be employed also in the infrared, simply substituting the known experimental masses (and in some cases residues, as for $f_{\pi}$ ) of mesons and glueballs, in order to get correlators and $S$-matrix amplitudes that are both factorized over poles of physical particles and are asymptotic to the correct result in the ultraviolet. While in this paper we outline briefly just a few examples of asymptotic correlators, relevant for the light by light scattering amplitude (see section \ref{s9}) \footnote{Light by light scattering enters the $QCD$ corrections to the anomalous magnetic moment of the muon.} and for the structure of the pion form factor, there are many more
other applications of physical relevance that we will not discuss here.  \par 
Now, to illustrate the structure of our main results, we report below the effective action whose quantization provides the asymptotic generating functional of the $S$-matrix amplitudes order by order in the $\frac{1}{N}$ expansion in the scalar sector with positive charge conjugation, i.e. 
the generating functional of those amplitudes such that all the external states are scalar glueballs with positive charge conjugation, or certain scalar mesons or gluinoballs \footnote{In pure $\mathcal{N}=1$ $SUSY$ $YM$ no purely gluonic operators may have a non-vanishing vacuum expectation value \cite{Shifman1}. The argument on the spectral representation of the scalar coefficient functions in the $OPE$ in section \ref{s11} assumes instead that the scalar condensates do not vanish generically. Thus applying the asymptotically-free bootstrap to the scalar sector in $\mathcal{N}=1$ $SUSY$ $YM$ must involve operators that are not purely gluonic, i.e. scalar composite operators that involve also fermion fields. For simplicity we refer generically to the single-particle states created by such operators as gluinoballs.
In the $QCD$ case as well, we refer to the single-particle states created by operators that involve fermion bilinears with possible gluonic insertions as mesons.}, depending on the particular realization of the spectrum generating algebra that satisfies the assumptions of the asymptotically-free bootstrap (see section \ref{s7}), and no higher-spin particle propagates in intermediate states. \par 
At lower orders it reads \footnote{To keep the notation simple we write often $dp$ instead of $d^4p$, and so on.}:
\bea \label{can10}
S= &&  \frac{1}{2!} \sum_n \int dq_1 dq_2 \delta(q_1+q_2) \Phi_n(q_1) (q_1^2+m_n^2) \Phi_n(q_2) \nonumber \\
&&+ \frac{C}{3! N^g} \int dq_1 dq_2 dq_3 \delta(q_1+q_2+q_3)        
   \int \sum_{n_1=1}^{\infty} \frac{   \rho_0^{-\frac{1}{2}}(m^2_{n_1})\Phi_{n_1}(q_2)}{p^2+m^{2}_{n_1}  } \nonumber\\
 &&\sum_{n_2=1}^{\infty}   \frac{ \rho_0^{-\frac{1}{2}}(m^{2}_{n_2})\Phi_{n_2}(q_3)}{(p+q_2)^2+m^{2}_{n_2}  }   
   \sum_{n_3=1}^{\infty}  \frac{ \rho_0^{-\frac{1}{2}}(m^2_{n_3})\Phi_{n_3}(q_1)}{(p+q_2+q_3)^2+m^{2}_{n_3}  }dp + \cdots \nonumber \\
\eea
with $g=1$ for glueballs and gluinoballs and $g=\frac{1}{2}$ for mesons in 't Hooft large-$N$ limit, where $\Phi_n(q)$ are the Fourier modes of a real local field $\Phi_n(x)$, with canonical kinetic term, that describes the $n$-th scalar glueball (or meson or gluinoball) that occurs in the corresponding spectrum generating algebra. The three-point scalar vertex is uniquely fixed asymptotically, up to the overall normalization $C$, by conformal symmetry at lowest order of perturbation theory, plus $RG$, plus the operator product expansion ($OPE$), plus the asymptotic theorem (see section \ref{s2}). \par 
The constant $C$ is on the order of $1$, and it can be fixed in $QCD$ in the scalar glueball (or meson) sector, and in $\mathcal{N}$ $=1$ $SUSY$ $YM$ in the scalar gluinoball sector, by means of lowest-order perturbation theory. \par
The generating functional of the asymptotic
correlators is much more complicated, because it contains explicit information, via the factors of $Z_n$, on the anomalous dimension of the operator $\mathcal{O}^{(0)}$ of naive dimension $D$ that occurs in the correlators. At lower orders it reads:
\bea \label{eff10}
\Gamma= &&  \frac{1}{2!} \sum_n \int dq_1 dq_2 \delta(q_1+q_2) m_n^{4-2D} Z_n^{-2}\rho_0(m^{2}_{n}) \Phi_n(q_1) (q_1^2+m_n^2) \Phi_n(q_2) \nonumber \\
&&+ \frac{C}{3! N^{g}} \int dq_1 dq_2 dq_3 \delta(q_1+q_2+q_3)        
   \int \sum_{n_1=1}^{\infty} m_{n_1}^2 \frac{m^{-D}_{n_1} Z^{-1}_{n_1} \Phi_{n_1}(q_2)}{p^2+m^{2}_{n_1}  } \nonumber\\
 &&\sum_{n_2=1}^{\infty}   m_{n_2}^2 \frac{m^{-D}_{n_2} Z^{-1}_{n_2}\Phi_{n_2}(q_3)}{(p+q_2)^2+m^{2}_{n_2}  }   
   \sum_{n_3=1}^{\infty}      m_{n_3}^2\frac{m^{-D}_{n_3} Z^{-1}_{n_3}\Phi_{n_3}(q_1)}{(p+q_2+q_3)^2+m^{2}_{n_3}  }dp + \cdots \nonumber \\
\eea
While the asymptotic structure of three-point correlators is essentially unique, because it is based on the essential uniqueness of the corresponding conformal limit, much less information can be obtained in general for higher-point correlators. 
However, the $OPE$ combined with the asymptotic theorem for certain coefficient functions in the $OPE$ implies that there exist primitive $r$-point vertices ($r > 3$), up to normalization, in the effective action at order $N^{g(2-r)}$ ($g=1$ for glueballs or gluinoballs, $g=\frac{1}{2}$ for mesons) with the following
structure of non-local one-loop graphs, that generalize the structure of the three-point vertex:
\bea \label{hl0}
 &&\int dq_1 dq_2 \cdots dq_r \delta(q_1+q_2+ \cdots + q_r)        
   \int \sum_{n_1=1}^{\infty} m_{n_1}^2 \frac{m^{-D}_{n_1} Z^{-1}_{n_1} \Phi_{n_1}(q_2)}{p^2+m^{2}_{n_1}  } 
 \sum_{n_2=1}^{\infty}   m_{n_2}^2 \frac{m^{-D}_{n_2} Z^{-1}_{n_2}\Phi_{n_2}(q_3)}{(p+q_2)^2+m^{2}_{n_2}  }   \nonumber \\
&&\cdots   \sum_{n_r=1}^{\infty}      m_{n_r}^2\frac{m^{-D}_{n_r} Z^{-1}_{n_r}\Phi_{n_r}(q_1)}{(p+q_2+ \cdots +q_r)^2+m^{2}_{n_r}  }dp
\eea
and analogously for the primitive multi-point vertices, up to normalization, in the generating functional of the $S$-matrix amplitudes:
\bea \label{hlS0}
&&\int dq_1 dq_2 \cdots dq_r \delta(q_1+q_2 + \cdots+q_r)        
  \int \sum_{n_1=1}^{\infty} \frac{   \rho_0^{-\frac{1}{2}}(m^2_{n_1})\Phi_{n_1}(q_2)}{p^2+m^{2}_{n_1}  } 
 \sum_{n_2=1}^{\infty}   \frac{ \rho_0^{-\frac{1}{2}}(m^{2}_{n_2})\Phi_{n_2}(q_3)}{(p+q_2)^2+m^{2}_{n_2}  }   \nonumber \\
&&   \cdots \sum_{n_r=1}^{\infty}  \frac{ \rho_0^{-\frac{1}{2}}(m^2_{n_r})\Phi_{n_r}(q_1)}{(p+q_2+\cdots+q_r)^2+m^{2}_{n_r}  }dp  \nonumber \\
\eea
These contributions are singled out because they lead to a leading asymptotic behavior in multi-point correlators and to primitive vertices in the effective action, i.e. to vertices that cannot be obtained gluing lower-order vertices and propagators in the effective action. But in general, despite they carry the leading asymptotic behavior compatible with the given anomalous dimension, they need not to be the only contributions with such leading asymptotic behavior for $r>3$. These vertices, up perhaps to normalization, coincide with the loop expansion of a functional determinant in the one-loop effective action of a $\Phi^3$ theory.  \par
In fact, in the pure scalar glueball sector of positive charge conjugation of large-$N$ $QCD$, taking advantage that the $OPE$ arises by an algebra of free bosonic fields in the conformal limit (see section \ref{s4}),
the aforementioned vertices can be resummed to all the $\frac{1}{N}$ orders to give the asymptotic generating functional of the $S$-matrix a very compact form, i.e. a kinetic term plus the logarithm of a functional determinant of Fredholm type:
 \bea \label{S0}
S = \frac{1}{2} tr \int   \Phi(-\Delta +  M^2)  \Phi  \, d^4x + \frac{\kappa}{2} N^{2} \log Det_3 (- \Delta +  M^2 -  \frac{c'} { N} \rho_0^{-\frac{1}{2}} \Phi)
\eea
with the constants $\kappa,c'$ fixed by lowest-order perturbation theory.\par
Some explanations are in order. \par
The field $\Phi(x)$ is a Hermitian diagonal infinite matrix whose real eigenvalues $\Phi_n(x)$ are the local fields associated to each scalar glueball. \par
The Fredholm determinant $Det_3$ is the functional determinant deprived of the first-two traces in the expansion of its logarithm.
The diagonal mass matrix $M$ with eigenvalues $m_n$, and the diagonal matrix of the scalar spectral distribution $\rho_0$, with eigenvalues $\rho_0(m_n^2)$ and with dimension of  $\Lambda_{QCD}^{-2}$, are undetermined by the methods of the asymptotically-free bootstrap, but for the fact that the eigenvalues of the mass matrix are a discrete diverging sequence by assumption. \par
It is easy to check that the expansion of Eq.(\ref{S0}) reproduces at lower orders Eq.(\ref{can10}), and at all higher orders Eq.(\ref{hlS0}) up to perhaps overall normalization of each term, provided the product of the matrix insertions $((- \Delta +  M^2 )^{-1}\rho_0^{-\frac{1}{2}} \Phi)$ in the expansion of the functional determinant is interpreted as a tensor product in the internal indices \footnote{The trace of a tensor product of matrices is the product of the traces.}:
\bea
&&\Tr_3 \log(1- \frac{c'} { N} ((- \Delta +  M^2 )^{-1}\rho_0^{-\frac{1}{2}} \Phi)) \nonumber\\
&&=\hat Tr_3 \log(1- \frac{c'} { N} tr((- \Delta +  M^2 )^{-1}\rho_0^{-\frac{1}{2}} \Phi))
\eea
where $tr$ is the trace on the internal indices and $\hat Tr_3$ is the trace on space-time with the first-two traces skipped. \par
In fact, the asymptotic structure implied by Eq.(\ref{can10}, \ref{eff10}, \ref{hl0}, \ref{hlS0}) and by Eq.(\ref{S0}), sets the strongest constraints on any solution of $QCD$, in particular on a string solution of $QCD$ (see section \ref{s1}):
The asymptotic interaction in the scalar sector is non-local and generated by an infinite number of primitive vertices that look like one-loop diagrams in a $\Phi^3$ field theory (but with an infinite number of fields) up perhaps to normalization, one for each order of the $\frac{1}{N}$ expansion. \par
Because of the explicit factor of $\rho_0^{-\frac{1}{2}}$ in Eq.(\ref{hlS0}) and Eq.(\ref{S0}), that has dimension of $\Lambda_{QCD}$, if we limit ourselves just to power counting, the $S$-matrix (see section \ref{s5}) in the scalar sector behaves in the $UV$ 
as in a super-renormalizable field theory (but with an infinite number of fields), as opposed to ultrasoft $UV$ behavior typical of conventional string theories in flat space-time. To be precise, our asymptotic effective action for the $S$-matrix in the scalar sector corresponds to a certain choice of the contact terms in the coefficient functions of the $OPE$ that determine the asymptotic scalar interaction (see comment below Eq.(\ref{2S})).
Indeed, an important point is that the two-point correlators of composite operators have ambiguities by contact terms (see Eq.(\ref{at1}) and Eq.(\ref{at2})). These ambiguities by contact terms extend to the coefficient functions of the $OPE$ (see section \ref{s11}) and therefore to the three-point asymptotic correlators as determined by the asymptotically-free bootstrap. However, our choice of contact terms is in a sense canonical, i.e. it corresponds to the Kallen-Lehmann representation associated to scalar local free massive fields (see section \ref{s3}). \par
Besides, the ultrasoft behavior in string theories, for example, of the Veneziano amplitude \cite{Veneziano} for scalar particles in flat space-time (see \cite{GSW} for a modern review of string theory) refers to the complete (tree) scalar amplitude, that involves a sum on all higher-spins in the intermediate states, while our statement about super-renormalizability refers to just partial amplitudes that involve only scalar states as intermediate states. For large-$N$ $QCD$ (and for $\mathcal{N}=1$ $SUSY$ $YM$) the actual $UV$ behavior of the asymptotic $S$-matrix  requires an analog resummation on all the intermediate states, that is outside the scope of this paper. \par
Moreover, the actual behavior of the asymptotic $S$-matrix amplitudes restricted to the scalar sector as a function of the masses of the asymptotic states is very sensitive, because of the factors of  $\rho_0^{-\frac{1}{2}}$ for each external line, to the rate of grow with the mass of the spectral distribution, that includes the multiplicities (see section \ref{s5}), while the asymptotic spectral distribution always cancels replacing the sum by the integral (see section \ref{s11}) on intermediate states in internal propagators that link the primitive vertices in the generating functional of the asymptotic $S$-matrix. \par
In string theories defined on space-time curved in some extra dimensions it might not be impossible to reproduce such asymptotic behavior,
but no presently known string theory, in particular based on the $AdS$ String/Gauge Fields correspondence, reproduces the asymptotics of correlators implied by Eq.(\ref{eff10}). This statement follows already from just the structure of the two-point correlators, i.e. from the asymptotic theorem rather than from the asymptotically-free bootstrap, and it is discussed in detail section \ref{s1}. \par
To avoid confusion we stress that our statement about super-renormalizability by power counting in the scalar sector refers to the aforementioned non-perturbative $\frac{1}{N}$ expansion, that from the point of view of perturbative $QCD$ has been already renormalized at the leading large-$N$ order to generate the non-perturbative scale $\Lambda_{QCD}$. \par
The asymptotic $S$-matrix generating functional for spin-$1$ mesons (and gluinoballs) restricted to the spin-$1$ sector is reported in section \ref{s9}.
As in the scalar sector, spin-$1$ vertices display at lowest $\frac{1}{N}$ order a cubic-like non-local interaction with the same structure of a one-loop graph, that persists in multi-point vertices,
but despite the one-loop structure super-renormalizability by power counting only is lost: In the spin-$1$ sector the large-$N$ theory looks renormalizable by power counting, but not super-renormalizable. For spin-$1$ there are ambiguities similar to the scalar case in the choice of contact terms, that again are resolved canonically by the Kallen-Lehman representation for free local massive spin-$1$ fields. \par 
But all the possible divergences in the asymptotic $\frac{1}{N}$ expansion of the generating functional of the $S$-matrix that have a physical meaning must be reabsorbed in a (subleading in $\frac{1}{N}$) redefinition of $\Lambda_{QCD}$, that it is the only parameter of the theory in the (asymptotic) $S$-matrix.\par
The plan of the paper is as follows. \par
In section \ref{s1} we further discuss the physical meaning and implications of the asymptotically-free bootstrap. \par
In section \ref{s2} we find out the unique asymptotic solution, up to overall normalization, in terms of glueball, meson and gluinoball propagators for two- and three- point correlators of scalar single-trace gauge-invariant operators or of operators involving fermion bilinears in 't Hooft large-$N$ limit, employing the aforementioned asymptotically-free bootstrap based on purely field-theoretical methods.   \par
The asymptotic solution is written in terms of the unknown spectral density for fixed spin and in terms of the anomalous dimension of the operators, thus extending to three-point scalar correlators
the aforementioned asymptotic structure theorem for two-point correlators \cite{MBN}. \par
In section \ref{s3} we extend by means of the $OPE$ the unique asymptotic solution for the two- and the three-point scalar correlators to certain contributions to the $r$-point scalar correlators. These contributions satisfy
asymptotically the appropriate Callan-Symanzik equation in $QCD$ (and in $\mathcal{N}$ $=1$ $SUSY$ $YM$), and they furnish a leading asymptotic contribution for large momentum to the exact correlators restricted to the scalar sector. \par
In section \ref{s4} we employ the aforementioned asymptotic correlators to write down an asymptotic effective action of massless large-$N$ $QCD$ (and to some extent of $\mathcal{N}=1$ $SUSY$ $YM$) in the scalar 
sector of positive charge conjugation to all the $\frac{1}{N}$ orders, up to the overall normalization of the vertices,
that is fixed either by lowest-order perturbation theory or by exploiting the one-loop integrability in the large-$N$ Ferretti-Heise-Zarembo sector \cite{Fe} of both $QCD$ and $\mathcal{N}=1$ $SUSY$ $YM$. \par
In section \ref{s5} the asymptotic effective action is used to construct the generating functional of the asymptotic $S$-matrix amplitudes in the aforementioned scalar sector. 
Remarkably, as a consequence of the $LSZ$ reduction formulae, the (asymptotic) $S$-matrix depends only on the unknown spectral densities but not on the naive and
anomalous dimensions. Indeed, the $S$-matrix must not depend on the choice of the interpolating field for any fixed asymptotic state. \par
As a consequence, part of the transcendental information \footnote{We use this term both in a mathematical sense, because of the link of the large-$N$ planar diagrams with type $II_1$ non-hyperfinite von Neumann algebras (see \cite{Top} and references therein), and in a loose sense.} implied by summing planar diagrams for the exact anomalous dimensions disappears from the (asymptotic) $S$-matrix. This fact has far-reaching consequences, because the asymptotic $S$-matrix depends only on the unknown spectral densities. This vast simplification opens the way for a new search of the $QCD$ string (and of the 
$\mathcal{N}=1$ $SUSY$ $YM$ string), limited only to the spectrum and the $S$-matrix, that we suggest in section \ref{s1}. \par
In section \ref{s9} we outline how our results extend explicitly from the scalar case to spin-$1$ correlators, and we mention briefly the relation with important physical applications, that will be considered in more detail elsewhere: Light by light scattering, charged and neutral pion form factors, and the associated vector dominance.\par
In sections \ref{s7}
to be concrete, we list briefly a collection of operators generating the spectrum to which the asymptotically-free bootstrap applies, for $QCD$ glueballs and mesons, and for $\mathcal{N}=1$ $SUSY$ $YM$ gluinoballs. \par
In section \ref{s11} we extend the asymptotic theorem to the coefficients of the $OPE$ in the scalar case, and we discuss to some extent the higher-spin case. \par

\section{Conclusions and outlook} \label{s1}

The main limitation of the asymptotic solution presented in this paper is that it does not provide spectral information. \par
But first and foremost, the intrinsic interest of the asymptotic solution is that it furnishes a concrete guide to find out an actual solution, possibly only for the spectrum and the $S$-matrix amplitudes, by other methods,
that may be of field-theoretical or of string-theoretical nature. \par
Besides, the asymptotic solution is an easy way to check any present or forthcoming proposed solution (in the sense of the $\frac{1}{N}$ expansion), since such a solution is likely to be based on some new mathematical insight, necessarily unfamiliar, while the asymptotic solution
is based on well-established field-theoretical methods.  \par
Moreover, for approximate solutions, the asymptotic solution provides a quantitative measure of how good or bad the approximation is. 
For example, employing the asymptotic theorem for two-point correlators \cite{MBN} or directly resumming the leading logarithms of perturbation theory \cite{MBM}, it has become apparent that all the present proposal for the scalar or pseudoscalar glueball propagators in confining asymptotically-free $QCD$-like theories based on the $AdS$ String/large-$N$ Gauge Theory correspondence disagree by powers of logarithms \cite{MBM,MBN} with the asymptotic solution. \par
Indeed, by the asymptotic theorem the asymptotic behavior of the scalar glueball propagator, the correlator that controls the mass gap in large-$N$ $YM$, reads in any asymptotically-free gauge theory massless in perturbation theory \footnote{This estimate follows from the anomalous dimension of $Tr{F}_{}^2$, that is $2 \beta_0$, that in turn follows from the $RG$ invariance of the conformal anomaly
$\frac{\beta(g)}{g} Tr{F}_{}^2$ (see for a detailed argument section 2.3 in \cite{MBM}, and \cite{MBM} for a perturbative check in massless $QCD$ as well).}:
\begin{align}\label{eqn:corr_scalare_inizio}
&\int\langle Tr{F_{}^2}(x) Tr{F}_{}^2(0)\rangle_{conn}e^{ip\cdot x}d^4x 
\sim  p^4\Biggl[\frac{1}{\beta_0\log\frac{p^2}{\Lambdams^2}}\Biggl(1-\frac{\beta_1}{\beta_0^2}\frac{\log\log\frac{p^2}{\Lambdams^2}}{\log\frac{p^2}{\Lambdams^2}}\Biggr)+O\biggl(\frac{1}{\log^2\plms}\biggr)\Biggr]
\end{align} 
up to contact terms, while all the scalar glueball correlators presently computed in the literature on the basis of the $AdS$ String/Gauge Fields correspondence behave as $p^4 \log^n (\frac{p^2}{\mu^2})$, with $n=1$ in the Hard Wall and Soft Wall models, and $n=3$ in the Klebanov-Strassler cascading $\mathcal{N}$ $=1$ $SUSY$ gauge theory, despite in the last case the asymptotically-free $NSVZ$ beta function is correctly reproduced in the supergravity approximation (see for details \cite{MBN,MBM} and references therein). \par
The aforementioned asymptotic disagreement implies that for an infinite number of poles and/or residues the large-$N$ glueball propagator on the string side of the would-be correspondence disagrees with the actual propagator of the asymptotically-free $QCD$-like theory on the gauge side. \par
This is unsurprising, since the stringy gravity side of the correspondence is in fact strongly coupled in the $UV$, and therefore it cannot describe the $UV$ of any confining asymptotically-free gauge theory. The asymptotic theorem furnishes a quantitative measure of how much in these models the asymptotic freedom is violated. \par
Thus there is no reason that the needed spectral information be correctly encoded in such class of strongly-coupled $AdS$-based theories \footnote{As a matter of fact these $AdS$ string-inspired models are currently applied to $QCD$ or to $QCD$-like asymptotically-free theories, as recalled in the recent vast review \cite{QCD}.}.\par
In fact, we should observe that in large-$N$ confining asymptotically-free gauge theories with no mass scale in perturbation theory, such as large-$N$ massless $QCD$ and $\mathcal{N}=1$ $SUSY$ $YM$, the non-perturbative dynamical generation of any physical mass scale
has not a strong coupling origin. \par
Indeed, the asymptotic theorem implies \cite{MBN} the following asymptotic structure of the scalar glueball propagator, since in this case $\gamma_0=2\beta_0$:
\bea 
\int \langle TrF^2(x) TrF^2(0) \rangle_{conn}\,e^{-ip\cdot x}d^4x  
\sim    \, p^{4}   \sum_{k=1}^{\infty} \frac{g^4(m_k^2)   \rho_0^{-1}(m^{2}_k)  }{p^2+m^{2}_k  }
\eea
Thus the residues of the poles of the scalar glueball propagator (deprived of the dimensionful factor of $p^4$, and of the factor of $\rho_0^{-1}$, that may be assumed to be constant asymptotically because of the expected asymptotic linearity of the spectrum of masses squared), in the far $UV$ vanish as $g^4(m^2_k)$ for large $k$, as required by the asymptotic freedom, while the scale $m_k^2$ of the corresponding poles diverges, in complete disagreement with the idea that the non-vanishing masses $m_k$ occur because the theory becomes strongly coupled: The more the theory becomes weakly coupled in the $UV$, the more massive poles occur in the $UV$, since they must be infinite in number to match at the leading large-$N$ order the $RG$-improved perturbative behavior, with vanishingly-small residues because of the weak coupling. \par
The real reason for the seemingly counter-intuitive behavior that any physical mass scale is dynamically generated at weak coupling is that any such physical scale must be proportional to the $RG$-invariant scale, that is constant along the whole $RG$ trajectory precisely because of its $RG$ invariance,
but in the continuum limit, i.e. for large cutoff, physical masses are much lower than the cutoff only at weak coupling. Since the mass spectrum necessarily must be a divergent sequence to match at the leading large-$N$ order
the $RG$-improved perturbative result, in order for the residues at the poles to contain the information about the beta function and the anomalous dimension in the deep ultraviolet, the cutoff must be necessarily sent to infinity and thus the coupling to zero. \par
On the contrary, if we insist that it makes sense to evaluate the mass gap at $g=\infty$ in any strong-coupling approach, then it coincides with the cutoff, because of the asymptotic freedom, since then $\Lambda_{QCD} \sim \Lambda e^{-\frac{1}{2 \beta_0 g^2}} \sim \Lambda$ (for more details see section 3 of \cite{MBFloer}), making the entire procedure of computing masses at $g=\infty$ in asymptotically-free theories meaningless, because they are on the same order of, or exceed, the poorly-defined cutoff.\par
Moreover, since it there exists a scheme in which the asymptotic theorem holds exactly \cite{MBN} and not only asymptotically, even for the lowest state, i.e. for the mass gap $m_1$, the coupling $g(m^2_1)$ cannot diverge and must occur in a scheme in which it is finite, for the Kallen-Lehmann representation to make sense. Thus even at the scale of the mass gap
the theory is not strongly coupled, in the sense that the coupling never diverges, as instead it does on the string side of the $AdS$ String/Gauge Theory correspondence in the (super-)gravity approximation. Thus there is really no reason for which even the low-lying mass ratios should come out reliable in that framework. \par
We discuss now the general meaning of the asymptotically-free bootstrap. \par
The $UV$ universality class represented by the asymptotically-free bootstrap for two- and three-point correlators is the non-trivial universality class defined by correlators of local gauge-invariant operators that makes sense
for confining large-$N$ asymptotically-free theories satisfying the assumptions in section \ref{s0}, with mass gap in the pure-glue sector, massless in perturbation theory. \par
On the contrary, the asymptotic behavior of correlators of local gauge-invariant operators in the infrared ($IR$) \footnote{For non-local observables such as Wilson loops there is a notion of large-$N$ universality class in the infrared defined by the Luscher term and 
its first subleading correction, that define the universality class of the large-$N$ $YM$ string in the infrared.} cannot define non-trivial universality classes \footnote{But in non pure-glue massless subsectors, if any.} for the aforementioned theories, since the $IR$ universality class is determined by the mass gap, that is the arbitrary free parameter, because it coincides with the $RG$-invariant scale in some scheme.\par
In fact, according to Witten \footnote{Talk at the Simons Center workshop: \emph{Mathematical Foundations of Quantum Field Theories}, Jan (2012), see in this respect \cite{Top}.}, any gauge theory with a mass gap defines a possibly-trivial topological field theory ($TFT$) in the $IR$. This is the much broader notion of universality in the $IR$ for theories with a mass gap, since it may contain trivial representatives of non-trivial theories.\par
Indeed, in the retrospect, we have proposed a $TFT$ underlying large-$N$ $YM$ constructed in \cite{MB0,MBJ,MBFloer} and analyzed in relation to the asymptotic theorem in \cite{MBN,MBM}, that realizes this idea in the appropriate technical sense.
The $TFT$ contains special topological Wilson loops invariant for deformations called twistor Wilson loops, that are completely trivial at the leading large-$N$ order (i.e. their v.e.v. is $1$), but non-trivial at next-to-leading order. But at next-to-leading order their non-triviality is controlled in large-$N$ $QCD$ by a certain anti-selfdual ($ASD$) propagator that is asymptotic \cite{MBN,MBM,MBFloer} to the asymptotic solution:
\bea \label{ASD}
\int \langle TrF^{-2}(x) TrF^{-2}(0) \rangle_{conn}\,e^{-ip\cdot x}d^4x  
=    \, \frac{2}{\pi^2} p^{4}   \sum_{k=1}^{\infty} \frac{g^4(k \Lambda^2_{QCD})   \Lambda^2_{QCD} }{p^2+k \Lambda^2_{QCD}  }
\eea
What makes the $TFT$ underlying $YM$ a powerful tool is the triviality of topological Wilson loops at all scales, in particular in the $UV$ and not only in the infrared, that allows us to localize them
on critical points by a quantum version of Morse-Smale-Floer homology \cite{MB0,MBJ,MBFloer}. But fluctuations around the $TFT$ at the next-to-leading $\frac{1}{N}$ order are non-topological, and give rise to the $ASD$ propagator \cite{MBN,MB0, MBFloer}, whose asymptotics defines a non-trivial universality class in the $UV$. \par
The philosophy of constructing a trivial topological sector underlying the large-$N$ limit of $YM$ or of $QCD$ is in sense a device that allow us to get a limited information on the spectrum and the local algebra, that is represented by free fields at the leading large-$N$ order, avoiding at the same time solving for the non-trivial Wilson loops, that contain already at the leading large-$N$ order an enormous amount of information \cite{Top}, not immediately relevant for the $QCD$ spectrum and the $S$-matrix. \par 
It is interesting to analyze how much information is contained in the various approaches to large-$N$ $QCD$, and therefore how much difficulty must be overcome: The complete large-$N$ solution that includes all the Wilson loops, 
the $TFT$ summarized in Eq.(\ref{ASD}), and the large-$N$ $S$-matrix. \par
In fact, a string solution in the modern sense contains exactly the same information as a complete solution of large-$N$ $QCD$ \cite{Top}.
Indeed, were we to know the $AdS$-inspired  String of $QCD$ at weak coupling, we would have mapped a difficult weak-coupling problem, the $QCD$ spectrum and correlators, into another difficult weak-coupling problem, the $AdS$-inspired String of $QCD$ at weak coupling, for which presently there is not even an explicit Lagrangian formulation, as opposed to the original field-theory problem for which we know the $QCD$ Lagrangian. \par
Moreover, what makes this weak-coupling problem intrinsically difficult is the modern philosophy of the large-$N$ exact String/Gauge fields duality, that requires exact equivalence at the level of all the Wilson loops and of all the local correlators, a vast amount of information in a technical sense, because the algebra of Wilson loops is never represented by free fields, not even at the leading large-$N$ order, as opposed to the two-point connected correlators of local operators, that are sums of propagators of free fields at the leading order. In fact, the ambient von Neumann algebra of all the Wilson loops, that results by summing all the planar diagrams at weak coupling, is 
 a non-hyperfinite von Neumann algebra, and as such it involves a vast information in a mathematical sense (see \cite{Top} and references therein).  \par
 Much less information is contained in the $TFT$, because of its triviality at the leading $\frac{1}{N}$ order. 
 Besides, at the next order Eq.(\ref{ASD}) involves just free fields, and contains information via a special correlator only on the joint scalar and pseudoscalar spectrum, and on the flow of coupling in a certain scheme. \par
Yet, the $TFT$ employs unconventional methods \cite{MBJ,MBFloer}, because the approach is field theoretical, and the most difficult part is to obtain massive fluctuations with masses proportional to $\Lambda_{QCD}$, a weak-coupling but highly non-perturbative problem. \par
Even less field-theoretical information is contained in the string approach to the $S$-matrix, that we advocate again here somehow more than forty years after the original Veneziano proposal \cite{Veneziano}. Indeed, in the light of the asymptotic solution only the spectral information matters for the (asymptotic) $S$-matrix (see section \ref{s5}).
Besides, the $QCD$ string that we look for should be by definition asymptotic to the asymptotic solution for the $S$-matrix, but should not contain any notion of the bare or running coupling, as opposed to the $AdS$-String approach, but only the string scale identified with $\Lambda_{QCD}$, not viewed as a function of the coupling yet. \par
In this way the field-theoretical problem, non-perturbative in the coupling, is completely avoided for the string $S$-matrix, but nevertheless the string candidate can be tested against the implications of the asymptotic freedom on the basis of the asymptotic solution for the $S$-matrix provided by the asymptotically-free bootstrap. \par
In this respect our philosophy aims to realize, according to Veneziano, the old-fashioned program of a string solution of $QCD$ based on obtaining the $S$-matrix rather than all the correlators, because of the vast simplification
that occurs in Eq.(\ref{can10}, \ref{hlS0}, \ref{S0}) due to the asymptotically-free bootstrap. Thus the approach that we advocate is, in a quite technical sense, much more modest than the modern philosophy of the Gauge Fields/Strings duality. \par
In fact, we propose conjecturally that the viable, i.e. solvable, string approach to the solution of $QCD$, for the spectrum and the (collinear) $S$-matrix only, is based on a certain Topological String Theory on (non-commutative) twistor space, essentially because it is dual to the aforementioned $TFT$ underlying the large-$N$ limit of $YM$ \cite{MBFloer}. This proposal will be explored in a forthcoming paper, but we would like to mention
why it arises in relation to the asymptotic solution in this paper.  \par
Perhaps the quickest way is by the following analogy, that requires a short detour. \par
According to \cite{WittenS}, the self-dual sector of $\mathcal{N}=4$ $SUSY$ $YM$ can be described at the string level by a Topological $B$-model on twistor space of ordinary space-time, whose effective action is holomorphic Chern-Simons on twistor space. In this sector the celebrated maximally elicity-violating amplitudes ($MHV$) of large-$N$  $YM$  \footnote{At tree level the $MHV$ amplitudes of large-$N$ pure $YM$ coincide with the ones of $\mathcal{N}=4$ $SUSY$ $YM$.} are realized as string amplitudes on twistor space. \par
According to \cite{MasonSkinner}, the twistorial formulation extends to a complete exact equivalence with $\mathcal{N}=4$ $SUSY$ $YM$ at the level of the functional integral,
provided the holomorphic Chern-Simons action is added to an interacting term, that is the logarithm of a certain functional determinant restricted to the fiber of twistor space. \par
More precisely, the exact equivalence is proved \cite{MasonSkinner} showing that there is a gauge in which the twistorial formulation reduces to the classical action of $\mathcal{N}=4$ $SUSY$ $YM$ in its standard space-time formulation,
and a gauge, not accessible by the space-time point of view, in which the logarithm of the aforementioned functional determinant in the twistorial formulation coincides with the generating functional of the tree
$MHV$ amplitudes. \par
The prescription of adding the logarithm of the functional determinant to the Chern-Simons action is therefore completely justified \cite{MasonSkinner} independently on the original string proposal \cite{WittenS}, that involves the same $MHV$ amplitudes but also other contributions in the effective action. \par 
The occurrence of the logarithm of the functional determinant, in the light of the asymptotic solution, is crucial also in our string proposal for $QCD$. \par
Indeed, formally the fact that the generating functional of the $S$-matrix in the self-dual sector of  $\mathcal{N}=4$ $SUSY$ $YM$ occurs as the logarithm of a functional determinant restricted to the fiber of twistor space is immediately reminiscent of the structure of the asymptotic $S$-matrix of large-$N$ $QCD$ presented in the previous section,
but this analogy cannot work as it stands, because $\mathcal{N}=4$ $SUSY$ $YM$ has zero beta function, it is exactly conformal, and it does not confine \footnote{In fact, suitably restricting the super-field
of the $\mathcal{N}=4$ theory, it can be shown that the twistorial formulation applies to theories with less supersymmetry, even to pure $YM$ \cite{MasonSkinner}. But this leads to the usual perturbative approach, that is of little interest from our 
non-perturbative point of view. }.\par
However, we build on this analogy in our search of the formulation appropriate for $QCD$. Indeed, already shortly after the original proposal in \cite{WittenS} it has been suggested in \cite{VN} that the $A$-version of the Topological String on twistor space
can extend to theories with less supersymmetry, in particular $\mathcal{N}=1$ and $\mathcal{N}=0$, essentially because the Topological $A$-model, as opposed to the $B$-model, does not require as consistency condition
to live on a  (super-)twistor space that is a (super-)Calabi-Yau manifold.\par 
The (twistorial) $A$-model lives on Lagrangian submanifolds. Hence we suggest that it may furnish the generating functional of the $S$-matrix restricted to only two components of the external momenta $(p_+, p_-)$, i.e. the collinear $S$-matrix
in the infinite momentum frame. \par
Besides, in addition to the Chern-Simons action restricted to Lagrangian submanifolds, an interaction term arises in the Topological $A$-model, by summing world-sheet instantons in the string formulation. With the usual combinatorics the interaction term arises by exponentiating Wilson loops wrapping the Lagrangian submanifolds \cite{VN}. \par
Yet the difficulty with the Topological $A$-model on twistor space, already pointed out in \cite{VN}, is that it does not seem to accommodate enough degrees of freedom to define particle states in its de Rham
cohomology, as opposed to the Dolbeault cohomology that occurs in the $B$-model, that provides massless particles thanks to the celebrated Penrose construction of the solution of massless wave equations by holomorphic
functions on twistor space. \par
However, we have already suggested some years ago \cite{MBstring} that the needed particle-like degrees of freedom can be embedded in the Topological $A$-model provided the space-time base of twistor space be transformed in the non-commutative manifolds $R^2 \times R^2_{\theta}$ or $R^2 \times T^2_{\theta}$ in the limit of infinite non-commutativity. \par
 Indeed, 
$YM$ on non-commutative space-time in the limit of infinite non-commutativity is equivalent to the large-$N$ limit of $YM$ on commutative space-time (see \cite{Top} for references).
Moreover, and more importantly, the non-commutativity allows us to encode the missing degrees of freedom in the twistorial $A$-model \cite{MBstring} in color degrees of freedom, by means of the modern version of the non-commutative Eguchi-Kawai reduction
or Morita equivalence, that allows us to trade space-time degrees of freedom into color and vice versa, essentially because translations in the non-commutative setting can be realized as gauge transformations (see for example \cite{Top} for a short review, and \cite{MBFloer} for technical details). \par
Yet at the time \cite{MBstring} we have not been able to construct a non-trivial $S$-matrix, essentially because we have employed in the twistorial $A$-model the usual combinatorics \cite{VN} for the interacting term that resums world-sheet instantons. \par
But we have reconsidered this problem in the light of the observation in \cite{MasonSkinner} that the $A$-model analog of the interaction term involving the logarithm of the functional determinant in the twistorial $B$-model can be recasted in a form
that involves an expansion into a sum of Wilson loop insertions, as usual but with a different combinatorics to account for the logarithm, that implies dividing by the number of cyclic permutations. \par
This leads to our proposal for the interacting term in the large-$N$ $YM$ string, that is the logarithm of a functional determinant restricted to the fiber, in entire analogy with the prescription in \cite{MasonSkinner} for the interacting term in holomorphic Chern-Simons and with the asymptotic $S$-matrix of large-$N$ $QCD$ in this paper.  \par
There is a final ingredient.
We pick Chern-Simons with non-Hermitian instead of Hermitian connection \cite{MBstring} (also known as Chern-Simons with complex gauge group). The reason is that the basic connection that gives rise to the anti-selfdual sector of $YM$ in the $TFT$ on the field-theory side \cite{MBFloer}, that contains the three $ASD$ components of the curvature, 
by the Ward construction in twistor space is not Hermitian \cite{MBstring}.  \par
Hence, following the analogy with \cite{MasonSkinner}, we would single out the twistorial $A$-model on non-commutative twistor space, eventually with some fermion degrees of freedom to include quark flavors, as the unique twistorial candidate for the large-$N$ $QCD$ string (collinear) $S$-matrix, 
by showing that there is a gauge in which the effective action reproduces the large-$N$ $QCD$ (one-loop) effective action around certain configurations, and a gauge, not accessible by the space-time point of view, in which we obtain a non-trivial $S$-matrix, that
is asymptotic to the field-theoretical asymptotic $S$-matrix in the $UV$, and it is factorized over glueball Regge trajectories linear in the masses squared, according to the $TFT$ \cite{MBFloer}. This is a program for the immediate future, that we will explore in a forthcoming paper. \par
To come back to purely field-theoretical arguments and to summarize, we may wonder what is the universal description of large-$N$ confining perturbatively-massless asymptotically-free gauge theories satisfying our assumptions, since by the aforementioned arguments it cannot be a theory evaluated at strong coupling or in the $IR$.  \par
As we already pointed out, in the $IR$ it there exists a reach-enough notion of universality only for truly massless theories, in terms of infrared behavior and critical exponents, but no such a notion exists for massive theories.
The asymptotic theorem for two-point correlators and the structure of the asymptotic three-point correlators show
that for large-$N$ confining asymptotically-free theories a non-trivial notion of universality
applies in fact to the $UV$ rather than to the $IR$, since it is determined by the anomalous dimensions in the $UV$, that are the analogs of the critical exponents in the $IR$.\par
The (string ?) solution of large-$N$ $QCD$ (and of large-$N$ $\mathcal{N}=1$ $SUSY$ $YM$) should be looked for
in the universality class explicitly constructed in the ultraviolet, rather than among theories that admit a mass gap, but are strongly coupled in the ultraviolet.  \par    

\section{Asymptotically-free bootstrap for massless large-$N$  $QCD$ and $\mathcal{N}=1$ $SUSY$ $YM$} \label{s2}

We start summarizing the basic ingredients of the asymptotically-free bootstrap, and then we describe the details. \par
The first step to work out the asymptotically-free bootstrap consists in exploiting the conformal invariance of two- and three-point scalar correlators, at the lowest non-trivial order of perturbation theory, together
with the $RG$ corrections implied by the Callan-Symanzik equation, in any asymptotically-free
theory massless in perturbation theory.\par
It follows the asymptotic estimate in Eq.(\ref{AS0}) for the three-point correlator of a scalar operator $\mathcal{O}$, that is proved in the first part of this section:
\bea \label{AS0}
&&G^{(3)}(\lambda(x_1-x_2), \lambda(x_2-x_3), \lambda(x_3-x_1))  \nonumber \\
&&\sim C(\lambda(x_1-x_2)) C(\lambda(x_2-x_3)) C(\lambda(x_3-x_1))+O(g^2(\lambda x_0))) \frac{Z^3(\lambda,g(\mu))}{\lambda^{3D}}
\eea
as $\lambda \rightarrow 0$, with the first term in the last line on the order of $\frac{Z^3(\lambda,g(\mu))}{\lambda^{3D}}$, i.e. leading, and $C(x)$ the coefficient function of $\mathcal{O}$ itself
in the $OPE$:
\bea \label{0}
\mathcal{O}(x)\mathcal{O}(0) \sim  C(x) \mathcal{O}(0) + \cdots
\eea
The basic idea is that Eq.(\ref{AS}) would be exact up to perhaps an overall constant in a conformal field theory, while it is only asymptotic in an asymptotically-free theory massless in perturbation theory. \par
Moreover, the key step of the asymptotically-free bootstrap, that applies specifically to the large-$N$ 't Hooft limit of confining asymptotically-free theories satisfying our assumptions \footnote{As explained in section \ref{s0}, we recall that by confining we mean gauge theories whose particle mass spectrum for fixed spin in large-$N$ 't Hooft limit
is a discrete divergent sequence at the leading large-$N$ order. We recall as well that by large-$N$ limit of 't Hooft type we mean a large-$N$ limit such that the connected two-point correlators of single-trace gauge invariant operators and of fermion bilinears are a sum of free propagators at the leading order, because the interaction and the particle widths
are suppressed at the leading order.}, occurs combining Eq.(\ref{AS0}) with Eq.(\ref{001}), that furnishes the asymptotic Kallen-Lehmann representation of $C(x)$ derived in section \ref{s11}:
\bea 
C(x_1-x_2)
&&\sim \sum_{n=1}^{\infty} \frac{1}{(2 \pi)^4} \int   \frac{m^{D-4}_n Z_n \rho^{-1}(m^{2}_n)}{p^2+m^{2}_n  } \,e^{ip\cdot (x_1-x_2)}dp \nonumber \\
&&\sim \sum_{n=1}^{\infty} \frac{1}{(2 \pi)^4} \int   \frac{m^{D-4}_n (\frac{g(m_n)}{g(\mu)})^{\frac{\gamma_0}{\beta_0}} \rho^{-1}(m^{2}_n)}{p^2+m^{2}_n  } \,e^{ip\cdot (x_1-x_2)}d^4p \nonumber \\
&& \sim  
\frac{(\frac{g(x_1-x_2)}{g(\mu)})^{\frac{\gamma_0}{\beta_0}}}{  (x_1- x_2)^{D}} 
\eea 
Thus combining Eq.(\ref{AS0}) with Eq.(\ref{001}) we get the spectral representation of the asymptotic interaction in Eq.(\ref{2}). \par
We define the two- and three-point connected correlators of a scalar operator $\mathcal{O}$ of naive dimension $D$:
\begin{equation}
\braket{\mathcal{O}(x_1)\mathcal{O}(x_2)}_{\mathit{conn}}=G^{(2)}(x_1-x_2)
\end{equation}
\begin{equation}
\braket{\mathcal{O}(x_1)\mathcal{O}(x_2) \mathcal{O}(x_3)}_{\mathit{conn}}= G^{(3)}(x_1-x_2, x_2-x_3, x_3-x_1)
\end{equation}
$G^{(2)}$ and $G^{(3)}$ satisfy the Callan-Symanzik equations:
\begin{equation} \label{CS2}
\left(\sum^{i=2}_{i=1} x_{i} \cdot \frac{\partial}{\partial x_{i}}+\beta(g)\frac{\partial}{\partial g}+2(D+\gamma(g))\right) G^{(2)}(x_1-x_2)  =0
\end{equation}
\begin{equation} \label{CS3}
\left(\sum^{i=3}_{i=1} x_{i} \cdot \frac{\partial}{\partial x_{i}}+\beta(g)\frac{\partial}{\partial g}+3(D+\gamma(g))\right) G^{(3)}(x_1-x_2, x_2-x_3, x_3-x_1)  =0
\end{equation}
At the lowest order they read:
\begin{equation}
\left(\sum^{i=2}_{i=1} x_{i} \cdot \frac{\partial}{\partial x_{i}} + 2 D\right) G^{(2)}(x_1-x_2)  =0
\end{equation}
\begin{equation}
\left(\sum^{i=3}_{i=1} x_{i} \cdot \frac{\partial}{\partial x_{i}} + 3 D \right) G^{(3)}(x_1-x_2, x_2-x_3, x_3-x_1)  =0
\end{equation}
The Callan-Symanzik equation exploits the action of the simultaneous rescaling of the coordinates, but at the lowest order, because of the vanishing of the beta function and of perturbative physical scales, the theory is actually conformal.
Exploiting conformal invariance, the actual solution at lowest order reads:
\begin{equation}\label{lo}
G^{(2)}(x_1-x_2)= C_2
\frac{1}{(x_1-x_2)^{2D}}
\end{equation}
\begin{equation}
G^{(3)}(x_1-x_2, x_2-x_3, x_3-x_1) = 
C_3 \frac{1}{(x_1-x_2)^{D}}\frac{1}{(x_2-x_3)^{D}}\frac{1}{(x_3-x_1)^{D}}
\end{equation}
with the constants $C_2,C_3$ computable at the lowest order of perturbation theory, and $x^D \equiv (x^2)^{\frac{D}{2}}$. Indeed, as it is well known, conformal invariance completely determines the structure of the two- and three-point scalar correlators up to overall normalization \footnote{To be precise, we assume that our scalar operators become conformal primary operators in the conformal limit.}, as opposed to just the scaling invariance implied by the
Callan-Symanzik equation at this order. We assume in this paper that $C_2,C_3$ are non-vanishing, as it occurs in the generic case. While $C_2$ is always non-vanishing in a unitary theory, $C_3$ may vanish in special cases, as discussed in section \ref{s7}. The non-vanishing of $C_3$ is a crucial assumption, and therefore the asymptotically-free bootstrap applies only to operators satisfying this condition. \par
At the order of $g^2$ the beta function is still zero, because it starts at order of $g^3$:
\bea
\frac{\partial g}{\partial \log \mu}=-\beta_0 g^3-\beta_1 g^5+\cdots
\eea 
and affects the correlators at the order of $g^4$, because of the derivative $\beta(g) \frac{\partial}{\partial{g}}$ in the Callan-Symanzik equation, but an anomalous dimension quadratic in the coupling arises in Eq.(\ref{CS2}) and Eq.(\ref{CS3}).
We get:
\begin{equation}
\left(\sum^{i=2}_{i=1} x_{i} \cdot \frac{\partial}{\partial x_{i}} + 2 (D-\gamma_0 g^2) \right) G^{(2)}(x_1-x_2)  =0
\end{equation}
\begin{equation}
\left(\sum^{i=3}_{i=1} x_{i} \cdot \frac{\partial}{\partial x_{i}} + 3 (D-\gamma_0 g^2) \right) G^{(3)}(x_1-x_2, x_2-x_3, x_3-x_1)  =0
\end{equation}
Since the beta function still vanishes at this order, the solution is still conformal, but with non-canonical dimensions. To the order quadratic in the coupling we get:
\bea \label{lo1}
&& G^{(2)}(x_1-x_2)= C_2(1+O(g^2(\mu)))
\frac{1}{(x_1-x_2)^{2 D-\gamma_0 g^2(\mu)  }} \nonumber \\
&&  \sim C_2(1+O(g^2(\mu)))
\frac{1}{(x_1-x_2)^{  2D  }}(1+\gamma_0 g^2(\mu) \log(|x_1-x_2|\mu))
\eea
\bea \label{lo3}
&&G^{(3)}(x_1-x_2, x_2-x_3, x_3-x_1) \sim C_3 (1+O(g^2(\mu)))
\frac{ (1+\gamma_0 g^2(\mu) \log(|x_1-x_2|\mu)) }{(x_1-x_2)^{D}} \nonumber \\
&&	\frac{ (1+\gamma_0 g^2(\mu) \log(|x_2-x_3|\mu)) }{(x_2-x_3)^{D}} 	\frac{ (1+\gamma_0 g^2(\mu) \log(|x_3-x_1|\mu)) }{(x_3-x_1)^{D}}
\eea
The important fact at this order is that the three-point correlator is still factorized in terms of two-point correlators, as a consequence of the conformal invariance persisting at this order. \par
Now at the next-to-next order a non-vanishing beta function arises, and the logarithms in the numerator can be expressed in terms of the running coupling 
by means of the fundamental asymptotic relation:
\bea \label{asy}
1+\gamma_0 g^2(\mu) \log(|x|\mu) \sim (1+\beta_0 g^2(\mu) \log(|x|\mu))^{\frac{\gamma_0}{\beta_0}} \sim (\frac{g(x)}{g(\mu)})^{\frac{\gamma_0}{\beta_0}}
\eea
implied again by the Callan-Symanzik equation, that resums the leading and next-to-leading logarithms in a way that is peculiar to the asymptotically-free theory, rather than to its lowest-order conformal limit.
Hence:
\begin{equation}\label{21}
 G^{(2)}(x_1-x_2) \sim  C_2 (1+O(g^2))
\frac{(\frac{g(x_1-x_2)}{g(\mu)})^{\frac{2\gamma_0}{\beta_0}}}{  (x_1- x_2)^{2D}} 
\end{equation}
\begin{equation} \label{31}
G^{(3)}(x_1-x_2, x_2-x_3, x_3-x_1)
 \sim
C_3 (1+O(g^2))\frac{(\frac{g(x_1-x_2)}{g(\mu)})^{\frac{\gamma_0}{\beta_0}}}{(x_1-x_2)^{D}}\frac{(\frac{g(x_2-x_3)}{g(\mu)})^{\frac{\gamma_0}{\beta_0}}}{(x_2-x_3)^{D}}\frac{(\frac{g(x_3-x_1)}{g(\mu)})^{\frac{\gamma_0}{\beta_0}}}{(x_3-x_1)^{D}}
\end{equation}
asymptotically within the universal, i.e. the scheme-independent, leading and next-to-leading logarithmic asymptotic accuracy:
\bea
g(x)^2 \sim
\frac{1}{\beta_0\log(\frac{1}{x^2 \Lambda^2_{QCD}})}\biggl(1-\frac{\beta_1}{\beta_0^2}\frac{\log\log(\frac{1}{x^2 \Lambda^2_{QCD}})}{\log(\frac{1}{x^2 \Lambda^2_{QCD}})}+O(\frac{1}{\log(\frac{1}{x^2 \Lambda^2_{QCD}})})    \biggr)
\eea
Indeed, Eq.(\ref{21}) and Eq.(\ref{31}) satisfy the appropriate Callan-Symanzik equations when all the coordinates are rescaled by a common vanishing factor $\lambda$, according to the general asymptotic solutions:
\bea\label{410}
&&G^{(2)}(g(\mu); \lambda(x_1-x_2)) \nonumber \\
&&\sim  (1+O(g^2(\lambda |x_1-x_2|))) \frac{(\frac{g(\lambda |x_1-x_2|)}{g(\mu)})^{\frac{2\gamma_0}{\beta_0}}}{\lambda^{2D}}
G^{(2)}(g(\lambda); x_1-x_2)
\eea
\bea\label{41}
&&G^{(3)}(g(\mu); \lambda(x_1-x_2), \lambda(x_2-x_3),\lambda( x_3-x_1)) \nonumber \\
&&\sim  (1+O(g^2(\lambda |x_0|))) \frac{(\frac{g(\lambda |x_0|)}{g(\mu)})^{\frac{3\gamma_0}{\beta_0}}}{\lambda^{3D}}
G^{(3)}(g(\lambda); x_1-x_2, x_2-x_3, x_3-x_1)
\eea
There is an important difference in the $1+O(g^2)$ corrections between Eq.(\ref{21}) and Eq.(\ref{31}).\par
In Eq.(\ref{21}) the factor $1+O(g^2)$ is in fact  $1+O(g^2(x_1-x_2))$, i.e. the correction to the leading asymptotic behavior is a function of only the running coupling at the scale
 $|x_1-x_2|$. \par
This can be easily seen from the general solution of Eq.(\ref{CS2}) (see section \ref{s11}):
 \begin{equation} \label{aleph}
G^{(2)}(x_1-x_2)= \frac{1}{(x_1-x_2)^{2D}}\,\mathcal{G}_{0}(g(x_1-x_2))\, Z^2((x_1-x_2) \mu,g(\mu)) 
\end{equation}
and from the estimate $\mathcal{G}_{0}(g(x)) \sim 1+O(g^2(x))$, due to the assumption that the two-point correlator does not vanish at the lowest order in perturbation theory.\par
In Eq.(\ref{31}) instead the $O(g^2)$ correction is a dimensionless function of the running coupling and of the coordinates, whose structure is unknown \footnote{ To the order of $g^2$ no explicit  coordinate dependence can arise in the correlator but through the running coupling $g(x)$, because of the factorization implied by conformal invariance in Eq.(\ref{lo3}).}, unless computed order by order in perturbation theory, but for the fact that it is on the order of $O(g^2(\lambda |x_0|))$
when all the coordinates are rescaled by a common vanishing factor $\lambda$, according to Eq.(\ref{41}). \par
However, the asymptotic estimate in the stronger factorized form of Eq.(\ref{31}) for the leading term \footnote{Only the leading term is factorized as displayed in Eq.(\ref{31}), the $O(g^2)$ corrections have the same scaling dimension $3D$, but need not to be factorized in that way.} when all the coordinates are rescaled by a common vanishing factor, can be proved by means of the $OPE$, that allows us to convey more local information than the Callan-Symanzik equation alone, and that essentially reduces the
asymptotic estimate for three-point correlators to estimates for two-point correlators and for certain coefficient functions in the $OPE$. \par 
This is a convenient way of getting estimates in the asymptoticaly-free theory because
both the two-point correlators and the coefficient functions in the $OPE$ satisfy Callan-Symanzik equations, whose general solutions have a structure as in Eq.(\ref{aleph}) and in Eq.(\ref{aleph1}). \par
Moreover, the important feature for the aims of the asymptotically-free bootstrap is to reinterpret, under the assumption that $C_3 \neq 0$, Eq.(\ref{31}) as:
\bea \label{AS}
G^{(3)}(x_1-x_2, x_2-x_3, x_3-x_1) \sim C(x_1-x_2) C(x_2-x_3) C(x_3-x_1)
\eea
within the leading and next-to-leading logarithmic asymptotic accuracy, where $C(x)$ is the coefficient function of the operator $\mathcal{O}$ itself in the $OPE$ of two factors of $\mathcal{O}$:
\bea \label{0}
\mathcal{O}(x)\mathcal{O}(0) \sim  C(x) \mathcal{O}(0) + \cdots
\eea
Indeed, under the assumption that the three-point correlator $\braket{\mathcal{O}(x)\mathcal{O}(0) \mathcal{O}(y)}_{\mathit{conn}}$ does not vanish at lowest order in perturbation theory, i.e. $C_3\neq 0$, we can substitute in the correlator, with asymptotic accuracy as $x$ vanishes, the contribution in the $OPE$ that contains the operator $\mathcal{O}$ itself.
For $C_3\neq 0$ this is the leading asymptotic contribution from the $OPE$ of two operator insertions of $\mathcal{O}$ to the three-point correlator of $\mathcal{O}$. In fact, in a conformal field theory
Eq.(\ref{AS}) would hold exactly up to perhaps a constant overall factor and not only asymptotically. \par
We show how this works in detail momentarily, but we first work out the consequences of Eq.(\ref{0}). \par
$C(x)$ satisfies again the Callan-Symanzik equation (see section \ref{s11}):
\begin{equation}
\left( x \cdot \frac{\partial}{\partial x}+\beta(g)\frac{\partial}{\partial g}+ D+\gamma(g)\right) C(x)  =0
\end{equation}
whose general solution is:
\begin{equation} \label{aleph1}
C(x)= \frac{1}{x^{D}}\,\mathcal{G}_{1}(g(x))\, Z(x \mu,g(\mu)) 
\end{equation}
with the $RG$-invariant function $\mathcal{G}_{1}(g(x))$ computable order by order in perturbation theory.\par
For further use we notice that the perturbative behavior of $\mathcal{G}_{0}(g(x))$ and  $\mathcal{G}_{1}(g(x))$ as a function of $g(\mu)$ can be employed to determine by which power of the running coupling $g(x)$ the expansion of $\mathcal{G}_{1}(g(x))$ and $\mathcal{G}_{0}(g(x))$ actually starts (see also section \ref{s11}). This plays an important role in determining which is the leading asymptotic contribution from the $OPE$ of two operator insertions of $\mathcal{O}$ to the three-point correlator of $\mathcal{O}$. \par
For example, we have already seen that the assumption $C_ 2\neq 0$ implies that $\mathcal{G}_{0}(g(x))$ must be on the order of $O(1)$ as $g(x) \rightarrow 0$ for $x \rightarrow 0$, since $\mathcal{G}_{0}(g(x))$ must be on the order of $O(1)$ as $g(\mu) \rightarrow 0$ in perturbation theory, because otherwise $G^{(2)}(x)$ would vanish at lowest order in perturbation theory contrary to the assumption, 
since $Z(x \mu,g(\mu))$ is always on
the order of $O(1)$ in the limit $g(\mu) \rightarrow 0$ in perturbation theory because of Eq.(\ref{asy}). \par
Similarly, under the assumption that $C_3 \neq 0$, $\mathcal{G}_{1}(g(x))$ must be  $O(1)$ as $g(\mu) \rightarrow 0$ in perturbation theory, because otherwise $C_3$ would vanish contrary to the assumption, since we show below that all the remaining contributions in the $OPE$ are at most on the order of
$O(g^2(\mu))$ in perturbation theory. Thus for $C_3 \neq 0$:
\bea \label{ab}
C(x) \sim \frac{(\frac{g(x)}{g(\mu)})^{\frac{\gamma_0}{\beta_0}}}{ x^{D}} 
\eea 
Hence, from Eq.(\ref{0}) and Eq.(\ref{21}), we get asymptotically for $x_1 \rightarrow x_2$:
\bea \label{OPE}
\braket{\mathcal{O}(x_1)\mathcal{O}(x_2) \mathcal{O}(x_3)}_{\mathit{conn}} &&\sim C(x_1-x_2)\braket{\mathcal{O}(x_2)\mathcal{O}(x_3)}_{\mathit{conn}} \nonumber \\
&& \sim C(x_1-x_2) G^{(2)}(x_2-x_3) \nonumber \\
&& \sim
C(x_1-x_2)C^2(x_2-x_3) \nonumber \\
&& \sim \frac{(\frac{g(x_1-x_2)}{g(\mu)})^{\frac{\gamma_0}{\beta_0}}}{ (x_1-x_2)^{D}}  \frac{(\frac{g(x_2-x_3)}{g(\mu)})^{\frac{2\gamma_0}{\beta_0}}}{ (x_2-x_3)^{2D}} 
\eea
that coincides with Eq.(\ref{31}). Therefore, because of the symmetric nature of Eq.(\ref{31}), by its symmetric perturbative expansion Eq.(\ref{lo3}), and by Eq.(\ref{OPE}), we get the fundamental result valid for $C_3 \neq 0$:
\bea 
G^{(3)}(x_1-x_2, x_2-x_3, x_3-x_1) \sim C(x_1-x_2) C(x_2-x_3) C(x_3-x_1)
\eea
within the leading and next-to-leading logarithmic asymptotic accuracy, the precise statement being:
\bea \label{AS0}
&&G^{(3)}(\lambda(x_1-x_2), \lambda(x_2-x_3), \lambda(x_3-x_1))  \nonumber \\
&&\sim C(\lambda(x_1-x_2)) C(\lambda(x_2-x_3)) C(\lambda(x_3-x_1))+O(g^2(\lambda x_0))) \frac{Z^3(\lambda,g(\mu))}{\lambda^{3D}}
\eea
as $\lambda \rightarrow 0$, with the first term in the last line on the order of $\frac{Z^3(\lambda,g(\mu))}{\lambda^{3D}}$, i.e. leading. We stress that the leading term is factorized as displayed, while the subleading corrections in general are not, but they are subleading when all the coordinates are rescaled uniformly. 
We now prove our previous statement about the $OPE$. \par
Denoting $\mathcal{O}$ as $\mathcal{O}_0$, the complete $OPE$ can be written schematically as:
\bea
\mathcal{O}_0(x)\mathcal{O}_0(0)&\sim&  C(x) \mathcal{O}_0(0) + \sum_{i \neq 0} C_{i}(x) \mathcal{O}_{i}(0) \nonumber \\
&+& \sum_{D' \neq D} C_{D'}(x) \mathcal{O}_{D'}(0)+\sum_{s \neq 0} C_{s}(x) \mathcal{O}^{(s)}(0) + \cdots
\eea
where the sum in the first line involves scalar operators with the same naive dimension of $\mathcal{O}$ and the sums in the second line involve scalar operators of different naive dimensions and higher-spin operators.\par
Since the theory admits a conformal limit at lower orders in perturbation theory, we assume that the operators explicitly displayed become primary operators in the conformal limit, while the dots represent operators that become their descendants (i.e derivatives of primary operators) in the conformal limit. \par
To the operators that become primary the following considerations apply. 
Firstly, we can ignore the contributions to the $OPE$ of all the higher-spin operators, since the two-point correlator with $\mathcal{O}$, $<\mathcal{O}^{(s)}(0) \mathcal{O}(y)>$, that results from inserting the $OPE$ in the three-point correlator, vanishes by Euclidean symmetry.  \par
Secondly, we consider the insertion of operators of naive dimension different from $\mathcal{O}$. \par
Employing again the Callan-Symanzik equation for $C_{D'}(x)$ and for $ <\mathcal{O}_{D'}(0) \mathcal{O}_0(y)>$, 
it follows that the contribution from the $OPE$ to the three-point correlator:
\bea
\sum_{D' \neq D} C_{D'}(x)  <\mathcal{O}_{D'}(0) \mathcal{O}_0(y)> 
\eea
behaves at most as:
\bea
 \sum_{D' \neq D} \frac{(\frac{g(x)}{g(\mu)})^{\frac{2\gamma_{0}-\gamma_{0D'}}{\beta_0}}}{ x^{2D-D'}} g^2(y) \frac{(\frac{g(y)}{g(\mu)})^{\frac{\gamma_{0D'}+\gamma_0}{\beta_0}}}{  y^{D'+D}} 
\eea
because the correlator $<\mathcal{O}_{D'}(0) \mathcal{O}(y)>$ is on the order at most $O(g^2(\mu))$ in perturbation theory, since two-point correlators of primary operators of different naive dimension vanish in the conformal limit. Thus this contribution of the $OPE$ is subleading at least by a factor of $g^2(\lambda y)$, when also $\lambda y \rightarrow 0$ together with $\lambda x \rightarrow 0$, with respect to 
$ C(\lambda x) <\mathcal{O}(0) \mathcal{O}(\lambda y)>$, as it follows from Eq.(\ref{OPE}), Eq.(\ref{ab}) and Eq.(\ref{21}). \par
Thirdly, there is the contribution of operators with the same naive dimension of $\mathcal{O}$, that may mix with $\mathcal{O}$ under renormalization: 
\bea
\sum_{i \neq 0} C_{i}(x) \mathcal{O}_{i}(0)
\eea
that needs separate consideration. \par
Again by the Callan-Symanzik equation, it behaves at most as:
\bea
 \sum_{i \neq 0} \frac{(\frac{g(x)}{g(\mu)})^{\frac{2\gamma_{0}-\gamma_{0i}}{\beta_0}}}{x^{D}} <\mathcal{O}_{i}(0) \mathcal{O}_0(y)>
\eea
Thus to complete our proof it suffices to prove that for $i \neq 0$, $<\mathcal{O}_{i}(0) \mathcal{O}_0(y)>$ is at most of order of $O(g^2(\mu))$ in perturbation theory. \par
We show that this is indeed the case, provided a suitable orthonormal conformal basis for the operators $\mathcal{O}_i$ is chosen. \par
At the lowest order of perturbation theory massless $QCD$ (and $\mathcal{N}=1$ $SUSY$ $YM$) are conformal free theories. Let $\mathcal{O}_i$ be the finite set of scalar (real) operators
of dimension $D$ that become conformal primary operators in the free limit. In this limit the two-point correlators in any basis read:
\bea
<\mathcal{O}_i(x) \mathcal{O}_k(0)>=  a_{i k} \frac{1}{x^{2D}}
\eea
The matrix $a_{i k}$ is a real Hermitian symmetric matrix and as such can be always diagonalized. Thus it exists always an orthonormal basis in the conformal limit:
\bea
<\mathcal{O}_i(x) \mathcal{O}_k(0)>=  \delta_{i k} \frac{1}{x^{2D}}
\eea
In particular:
\bea
<\mathcal{O}_i(x) \mathcal{O}_0(0)>= 0 
\eea
for $i \neq 0$ at $g(\mu)=0$. At order of $g^2(\mu)$ renormalization and operator mixing may arise, but:
\bea
<\mathcal{O}_i(x) \mathcal{O}_0(0)> \sim O(g^2(\mu))
\eea
for $i \neq 0$ in perturbation theory, otherwise it would contradict the orthonormality of the conformal basis at $g(\mu)=0$. Therefore:
\bea
<\mathcal{O}_i(x) \mathcal{O}_0(0)> \sim O(g^2(x))
\eea
for $i \neq 0$.\par
Finally, there is the contribution of the descendants represented by the dots. As $\lambda x \rightarrow 0$ the contribution of the descendants 
to $G^{(3)}$ is suppressed either because the two-point function with $\mathcal{O}(\lambda y)$ vanishes in the conformal limit because they are descendants of operators orthogonal to $\mathcal{O}$ in the orthonormal conformal basis of primary operators, or because being descendants of $\mathcal{O}$ itself in the conformal limit they have necessarily higher naive dimension than $\mathcal{O}$, and thus their coefficient functions are suppressed by dimensional analysis as $\lambda x \rightarrow 0$. Hence Eq.(\ref{OPE}) follows for $x_1 \rightarrow x_2$. \par
Moreover, for $x_1-x_2$ fixed, the contribution of the descendants $\mathcal{O}^{(j)}$ of $\mathcal{O}$ in the conformal limit is to transform the term $C(x_1-x_2)G^{(2)}(x_2-x_3)$ that arises from the $OPE$ in Eq.(\ref{0}) into
$C(x_1-x_2)C(x_2-x_3)C(x_3-x_1)$ in order to guarantee the factorization of the conformal three-point correlator, since all the other contributions from the $OPE$ to the three-point correlator vanish in the conformal limit. \par
Incidentally, this shows also how the descendants $\mathcal{O}^{(j)}$, that involve in the $OPE$ the two-point correlators $<\mathcal{O}^{(j)}(0) \mathcal{O}(\lambda y)>$ and their coefficient functions $C^{(j)}(\lambda x) \sim C^{(j_0)}(\lambda x) Z^2(\lambda x \mu)Z^{(j) -1}(\lambda x \mu)\mathcal{G}_j(g(\lambda x))$ with $C^{(j_0)}(x)$ the coefficient function in the conformal limit, conspire to recover the asymptotic factorization of $G^{(3)}$ in Eq.(\ref{41}) for $\lambda \rightarrow 0$ and $x_1=x,x_2=0,x_3=y$ fixed. Hence our proof is completed.   \par
So far so good. Everything that we have discussed is well known, but perhaps the asymptotic estimate Eq.(\ref{AS}) in any asymptotically-free gauge theory massless in perturbation theory.\par
Now it comes the interesting part. 
The asymptotic theorem extends to the coefficients of the $OPE$ in the scalar case (see section \ref{s11}), because they arise from the non-perturbative part, that involves condensates \footnote{We assume that the scalar condensates do not vanish generically. See footnote $7$ for an exception in $\mathcal{N}=1$ $SUSY$ $YM$.}, of the scalar two-point correlator, which the
Kallen-Lehmann representation applies to as well \footnote{There are subtleties associated to power-like suppressed corrections arising by the terms involving the Bernoulli numbers in the Euler-MacLaurin formula, that are discussed
in section \ref{s11}.}. In particular for $C(x)$ we get \footnote{There is an important caveat discussed below Eq.(\ref{osc}).}:
\bea \label{001}
C(x_1-x_2)
&&\sim \sum_{n=1}^{\infty} \frac{1}{(2 \pi)^4} \int   \frac{m^{D-4}_n Z_n \rho^{-1}(m^{2}_n)}{p^2+m^{2}_n  } \,e^{ip\cdot (x_1-x_2)}dp \nonumber \\
&&\sim \sum_{n=1}^{\infty} \frac{1}{(2 \pi)^4} \int   \frac{m^{D-4}_n (\frac{g(m_n)}{g(\mu)})^{\frac{\gamma_0}{\beta_0}} \rho^{-1}(m^{2}_n)}{p^2+m^{2}_n  } \,e^{ip\cdot (x_1-x_2)}d^4p \nonumber \\
&& \sim  
\frac{(\frac{g(x_1-x_2)}{g(\mu)})^{\frac{\gamma_0}{\beta_0}}}{  (x_1- x_2)^{D}} 
\eea
We can consider this relation an asymptotic refinement of the Kallen-Lehmann representation, and we will often refer to this relation in this paper as such refinement. \par
The basic idea of the asymptotically-free bootstrap is to substitute the refined Kallen-Lehmann representation for $C(x)$, Eq.(\ref{001}), in Eq.(\ref{AS}). Thus explicitly and constructively we get:
\bea
&& \braket{\mathcal{O}_{D, \gamma_0}(x_1)\mathcal{O}_{D, \gamma_0}(x_2) \mathcal{O}_{D, \gamma_0}(x_3)}_{\mathit{conn}} \sim \nonumber \\
&&\sum_{n_1=1}^{\infty} \frac{1}{(2 \pi)^4} \int   \frac{m^{D-4}_{n_1} Z_{n_1}\rho^{-1}(m^{2}_{n_1})}{p_1^2+m^{2}_{n_1}  } \,e^{ip_1\cdot (x_1-x_2)}dp_1 \nonumber \\
&&\sum_{n_2=1}^{\infty} \frac{1}{(2 \pi)^4} \int   \frac{m^{D-4}_{n_2} Z_{n_2}\rho^{-1}(m^{2}_{n_2})}{p_2^2+m^{2}_{n_2}  } \,e^{ip_2\cdot (x_2-x_3)}dp_2 \nonumber \\
&&\sum_{n_3=1}^{\infty} \frac{1}{(2 \pi)^4} \int   \frac{m^{D-4}_{n_3} Z_{n_3}\rho^{-1}(m^{2}_{n_3})}{p_3^2+m^{2}_{n_3}  } \,e^{ip_3\cdot (x_3-x_1)}dp_3 
\eea
The corresponding Fourier transform reads:
\bea
&& \braket{\mathcal{O}_{D, \gamma_0}(q_1)\mathcal{O}_{D, \gamma_0}(q_2) \mathcal{O}_{D, \gamma_0}(q_3)}_{\mathit{conn}}  \nonumber \\
&& \sim \sum_{n_1=1}^{\infty} \frac{1}{(2 \pi)^4} \int   \frac{m^{D-4}_{n_1} Z_{n_1}\rho^{-1}(m^{2}_{n_1})}{p_1^2+m^{2}_{n_1}  } \,e^{ip_1\cdot (x_1-x_2)} \,e^{-iq_1\cdot x_1}dx_1 dp_1 \nonumber \\
&&\sum_{n_2=1}^{\infty} \frac{1}{(2 \pi)^4} \int   \frac{m^{D-4}_{n_2} Z_{n_2}\rho^{-1}(m^{2}_{n_2})}{p_2^2+m^{2}_{n_2}  } \,e^{ip_2\cdot (x_2-x_3)} \,e^{-iq_2\cdot x_2}dx_2 dp_2     \nonumber \\
&&\sum_{n_3=1}^{\infty} \frac{1}{(2 \pi)^4} \int   \frac{m^{D-4}_{n_3} Z_{n_3}\rho^{-1}(m^{2}_{n_3})}{p_3^2+m^{2}_{n_3}  } \,e^{ip_3\cdot (x_3-x_1)} \,e^{-iq_3\cdot x_3}dx_3 dp_3  \nonumber \\
&&=\sum_{n_1=1}^{\infty}  \int   \frac{m^{D-4}_{n_1} Z_{n_1}\rho^{-1}(m^{2}_{n_1})}{p_1^2+m^{2}_{n_1}  } \delta(p_1-p_3-q_1) dp_1 \nonumber \\
&&\sum_{n_2=1}^{\infty}  \int   \frac{m^{D-4}_{n_2} Z_{n_2}\rho^{-1}(m^{2}_{n_2})}{p_2^2+m^{2}_{n_2}  } \delta(p_2-p_1-q_2) dp_2     \nonumber \\
&&\sum_{n_3=1}^{\infty}  \int   \frac{m^{D-4}_{n_3} Z_{n_3}\rho^{-1}(m^{2}_{n_3})}{p_3^2+m^{2}_{n_3}  } \delta(p_3-p_2-q_3) dp_3  
\eea
Hence we get the asymptotic solution for two-point:
\bea \label{002}
&&\langle \mathcal{O}(q_1) \mathcal{O}(q_2) \rangle_{conn}   \sim  \delta(q_1+q_2) \sum_{n=1}^{\infty}   \frac{m^{2D-4}_n Z_n^{2} \rho^{-1}(m^{2}_n)}{q_1^2+m^{2}_n  }
 \eea
and three-point scalar correlators:
\bea \label{2}
&& \braket{\mathcal{O}_{D, \gamma_0}(q_1)\mathcal{O}_{D, \gamma_0}(q_2) \mathcal{O}_{D, \gamma_0}(q_3)}_{\mathit{conn}}  \nonumber \\
&& \sim \delta(q_1+q_2+q_3) \sum_{n_1=1}^{\infty}       \sum_{n_2=1}^{\infty}    \sum_{n_3=1}^{\infty}      
   \int   \frac{m^{D-4}_{n_1} Z_{n_1}\rho^{-1}(m^{2}_{n_1})}{p^2+m^{2}_{n_1}  } 
   \frac{m^{D-4}_{n_2} Z_{n_2}\rho^{-1}(m^{2}_{n_2})}{(p+q_2)^2+m^{2}_{n_2}  }   
  \frac{m^{D-4}_{n_3} Z_{n_3}\rho^{-1}(m^{2}_{n_3})}{(p+q_2+q_3)^2+m^{2}_{n_3}  }dp \nonumber \\
\eea
Thus the three-point scalar correlator is asymptotic to a non-local three-point vertex that has the structure of an amputated one-loop vertex constructed by means of scalar glueball (or gluinoball or meson) propagators and a certain local scalar interaction vertex. This vertex has an ambiguity by the choice of contact terms that affect $C(x)$ in the coordinate representation (see section \ref{s11}). They are entirely analog to the ambiguities by contact terms that affect the two-point correlator (see section \ref{s11}). For example $C(x)$ can be represented as well as:
\bea \label{2S}
C(x_1-x_2)\sim \sum_{n=1}^{\infty} \frac{1}{(2 \pi)^4} \int   \frac{p^{D-4} Z_n \rho^{-1}(m^{2}_n)}{p^2+m^{2}_n  } \,e^{ip\cdot (x_1-x_2)}dp \nonumber \\
\eea
that differs by our previous choice just by contact terms, but this choice leads to a non-renormalizable generating functional of the asymptotic $S$ matrix in the scalar sector (see section \ref{s5}), while the choice in Eq.(\ref{2}) leads to a super-renormalizable scalar interaction in the asymptotic $S$-matrix. Similar ambiguities occur for higher-spin vertices. \par
But this cannot be the whole story.  \par
Indeed, while the three-point correlator that we wrote in Eq.(\ref{2}) is certainly asymptotic \footnote{Up to the contact terms that occur in the refined Kallen-Lehmann representation for $C(x)$.} to the correct result in $RG$-improved perturbation theory, it has not the correct pole structure.
\par
In fact, the occurrence of poles for on-shell external momenta is a consequence of the $OPE$ for large $q_2$:
\bea \label{ope}
&&  \braket{ \mathcal{O}_{D, \gamma_0}(q_1) \mathcal{O}_{D, \gamma_0}(q_2)\mathcal{O}_{D, \gamma_0}(q_3) }_{\mathit{conn}}  \nonumber \\
&& \sim  \delta(q_1+q_2+q_3)        
   C(q_2)  G^{(2)}(q_3)    \nonumber \\ 
&&  \sim \delta(q_1+q_2+q_3) \sum_{n_1=1}^{\infty}            
     \frac{m^{D-4}_{n_1} Z_{n_1}\rho^{-1}(m^{2}_{n_1})}{q_2^2+m^{2}_{n_1}  } \sum_{n_2=1}^{\infty}     \frac{m^{2D-4}_{n_2} Z^2_{n_2}\rho^{-1}(m^{2}_{n_2})}{q_3^2+m^{2}_{n_2}  }  \nonumber\\
     \eea
as it follows by Fourier transforming the second line in Eq.(\ref{OPE})
and substituting the refined Kallen-Lehmann representation. \par
On the contrary, Eq.(\ref{2}) has the structure of a one-loop amputated Feynman diagram, while it should instead carry on-shell poles for external momenta in Minkowski space-time due to the aforementioned poles. As a consequence the three-point $S$-matrix amplitude, implied by the correlator in Eq.(\ref{2}) applying the $LSZ$ reduction formulae, vanishes identically, because the product of the three on-shell zeroes due to the three inverse propagators in the $LSZ$ formulae
has at most one pole to cancel out due to the explicit form of the three-point scalar vertex in Eq.(\ref{2}), that is a one-loop Passarino-Veltman scalar diagram (see for example \cite{PV}).   \par
Moreover, the on-shell poles in the external momenta in Eq.(\ref{ope})
necessarily arise by attaching to a three-point vertex the three propagators that must occur by postulating the existence of an effective action expanded order by order in $\frac{1}{N}$. \par
Thus while Eq.(\ref{ope}) and Eq.(\ref{2}) have the same large-momentum asymptotics for simultaneous rescaling by $\lambda^{-1} \rightarrow \infty$ of the Euclidean momenta, the symmetric form and the $OPE$ form factorize on different cuts and poles in Minkwoski space-time. \par
The explanation is that the $RG$-improved perturbative three-point correlator is only asymptotic in the ultraviolet, and therefore it has an ambiguity by multiplicative functions of the external Euclidean momenta that are asymptotic to $1$ in the ultraviolet.
We may wonder as to whether there is a way to reconcile the two forms of the asymptotic three-point correlator, without destroying the correct Euclidean asymptotic behavior. In fact, the answer is affirmative and unique, modulo the cyclic ordering of the external momenta. \par
Hence we fix asymptotically the aforementioned ambiguity by requiring that our new improved three-point correlator carries a simple pole for each external momentum $(q_1,q_2,q_3)$ on shell in Minkowski space-time, in order for the three-point asymptotic $S$-matrix amplitude to be non-zero, but at the same time without changing the Euclidean asymptotic behavior of the correlator. \par
Hence the real structure of the Euclidean correlator must be asymptotically:
\bea \label{3}
&& \braket{\mathcal{O}_{D, \gamma_0}(q_1)\mathcal{O}_{D, \gamma_0}(q_2) \mathcal{O}_{D, \gamma_0}(q_3)}_{\mathit{conn}}  \nonumber \\
&& \sim  \delta(q_1+q_2+q_3)        
   \int \sum_{n_1=1}^{\infty}  \frac{m^{D-4}_{n_1} Z_{n_1}\rho^{-1}(m^{2}_{n_1})}{p^2+m^{2}_{n_1}  } \frac{m_{n_1}^2}{q_2^2+m_{n_1}^2}\nonumber\\
 &&\sum_{n_2=1}^{\infty}    \frac{m^{D-4}_{n_2} Z_{n_2}\rho^{-1}(m^{2}_{n_2})}{(p+q_2)^2+m^{2}_{n_2}  }   \frac{m_{n_2}^2}{q_3^2+m_{n_2}^2}
   \sum_{n_3=1}^{\infty}     \frac{m^{D-4}_{n_3} Z_{n_3}\rho^{-1}(m^{2}_{n_3})}{(p+q_2+q_3)^2+m^{2}_{n_3}  }   \frac{m_{n_3}^2}{q_1^2+m_{n_3}^2} dp \nonumber \\
\eea
where we used $\lim_{n \rightarrow \infty} \frac{m_n^2}{q^2+m_n^2}=1$. \par
The improved correlator thus obtained agrees now with the asymptotics, and with the pole structure as well, implied by the $OPE$. \par
Indeed, it agrees almost tautologically with the first line in Eq.(\ref{ope}) for large Euclidean $q_2^2$ and $q^2_3$, and secondly satisfies the much stronger requirement that it factorizes asymptotically on the same poles and residues in the first sum in the second line of Eq.(\ref{ope}) in Minkowski space-time. \par
To check this property, we evaluate 
 asymptotically employing freely the limit $\lim_{n \rightarrow \infty} \frac{m_n^2}{q^2+m_n^2}=1$ inside the integrals:
\bea \label{4}
&& \braket{\mathcal{O}_{D, \gamma_0}(q_1)\mathcal{O}_{D, \gamma_0}(q_2) \mathcal{O}_{D, \gamma_0}(q_3)}_{\mathit{conn}}  \nonumber \\
&& \sim  \delta(q_1+q_2+q_3)        
   \sum_{n_1=1}^{\infty} \frac{ m^{D-4}_{n_1} Z_{n_1}\rho^{-1}(m^{2}_{n_1})    }{q_2^2+m_{n_1}^2} \int  \sum_{n_3=1}^{\infty}  \frac{   m^{D-4}_{n_3}    Z_{n_3}\rho^{-1}(m^{2}_{n_3})  }{(p-q_1)^2+m_{n_3}^2}    \sum_{n_2=1}^{\infty}   \frac{ m^{D-4}_{n_2} Z_{n_2}\rho^{-1}(m^{2}_{n_2})  }{(p+q_2)^2+m^{2}_{n_2}  }   dp \nonumber \\
&& \sim  \delta(q_1+q_2+q_3)        
   \sum_{n_1=1}^{\infty} \frac{ m^{D-4}_{n_1} Z_{n_1}\rho^{-1}(m^{2}_{n_1})    }{q_2^2+m_{n_1}^2}     \int  \sum_{n_3=1}^{\infty}  \frac{   m^{D-4}_{n_3}    Z_{n_3}\rho^{-1}(m^{2}_{n_3})  }{p^2+m_{n_3}^2}    
  \sum_{n_2=1}^{\infty}   \frac{ m^{D-4}_{n_2} Z_{n_2}\rho^{-1}(m^{2}_{n_2})  }{(p-q_3)^2+m^{2}_{n_2}  }   dp \nonumber \\
&& \sim  \delta(q_1+q_2+q_3)        
   \sum_{n_1=1}^{\infty} \frac{ m^{D-4}_{n_1} Z_{n_1}\rho^{-1}(m^{2}_{n_1})    }{q_2^2+m_{n_1}^2}     \int C^2(x) e^{- i q_3 \cdot x}   dx \nonumber \\
&& \sim  \delta(q_1+q_2+q_3)        
   \sum_{n_1=1}^{\infty} \frac{ m^{D-4}_{n_1} Z_{n_1}\rho^{-1}(m^{2}_{n_1})    }{q_2^2+m_{n_1}^2}  G^{(2)}(q_3) \nonumber \\
\eea
that coincides asymptotically for Euclidean and for Minkowskian $q_2^2$ with the first sum in Eq.(\ref{ope}), and agrees with the second sum as well but only asymptotically for large $q_3^2$, because of the convolution in the third line above, since $C^2(x) \sim G^{(2)}(x)$ asymptotically for small $x$. \par
Thus the scalar correlator, for any large-$N$ confining asymptotically-free theory massless in perturbation theory satisfying our assumptions, is asymptotically:
\bea \label{3}
&& \braket{\mathcal{O}_{D, \gamma_0}(q_1)\mathcal{O}_{D, \gamma_0}(q_2) \mathcal{O}_{D, \gamma_0}(q_3)}_{\mathit{conn}}  \nonumber \\
&& \sim  \delta(q_1+q_2+q_3)        
   \int \sum_{n_1=1}^{\infty}  \frac{m^{D-4}_{n_1} Z_{n_1}\rho^{-1}(m^{2}_{n_1})}{p^2+m^{2}_{n_1}  } \frac{m_{n_1}^2}{q_2^2+m_{n_1}^2} \nonumber\\
 &&\sum_{n_2=1}^{\infty}    \frac{m^{D-4}_{n_2} Z_{n_2}\rho^{-1}(m^{2}_{n_2})}{(p+q_2)^2+m^{2}_{n_2}  }   \frac{m_{n_2}^2}{q_3^2+m_{n_2}^2}
   \sum_{n_3=1}^{\infty}     \frac{m^{D-4}_{n_3} Z_{n_3}\rho^{-1}(m^{2}_{n_3})}{(p+q_2+q_3)^2+m^{2}_{n_3}  }   \frac{m_{n_3}^2}{q_1^2+m_{n_3}^2} dp \nonumber \\
\eea
This is a key result in this paper. \par
We may wonder why only scalar states circulate in the asymptotic scalar three-point vertex. In fact, this is a feature of 't Hooft large-$N$ limit at leading order. Indeed,
inserting a complete set of states in the operator version of the correlator (that involves a $T$ product) we get:
\bea
\sum_{n,m} <0|\mathcal{O}_{D, \gamma_0}(q_1)|n><n|\mathcal{O}_{D, \gamma_0}(q_2)|m><m| \mathcal{O}_{D, \gamma_0}(q_3)|0>
\eea
and the statement follows, since $\mathcal{O}$ can create from the vacuum only one-particle scalar states at leading order in 't Hooft large-$N$ limit. It would not be true in general in Veneziano large-$N$ limit.\par
We believe that analog arguments hold for higher-spin three-point correlators of sufficiently high dimension with respect to the twist (i.e. dimension minus spin), but in this paper we work out in detail in section \ref{s9} only the spin-$1$ case for conserved currents of naive dimension $3$ and twist $2$ constructed by fermion bilinears. It turns out that instead the latter case is somehow special, because a certain asymptotic estimate in section \ref{s11} becomes borderline.  

\section{Extension of the asymptotically-free bootstrap to primitive $r$-point scalar vertices} \label{s3}

If we restrict to the scalar sector of positive charge conjugation, we can get some information on the asymptotic structure of multi-point correlators involving only scalar states in the external lines and in the internal
propagators, by applying repeatedly the $OPE$, and taking into account that at the lowest order the $OPE$ arises by an algebra of free fields.\par
We employ:
\bea \label{OPE1}
\mathcal{O}(x_1)\mathcal{O}(x_2) \sim  C(x_1-x_2) \mathcal{O}(x_2) + \cdots
\eea
as a leading asymptotic contribution in a correlator with multiple insertions of $\mathcal{O}$ from the $OPE$ in the scalar sector. The proof consists in checking that Eq.(\ref{OPE1}) applied inductively below leads to a result that is asymptotic to the one implied by the Callan-Symanzik equation, when all the coordinates are rescaled by a common vanishing factor, because each factor of $C(x_i-x_j)$ contributes asymptotically, up to normalization, a factor of $Z((x_i-x_j)\mu, g(\mu)) \frac{1}{(x_i-x_j)^D}$
in the correlator, and no extra factor of the running coupling arises, thus implying the largest asymptotic behavior, when all the coordinates are rescaled by the common factor of $\lambda \rightarrow 0$, compatible with the anomalous dimension of $\mathcal{O}$. We stress that, contrary to the case of the three-point correlator, the argument above shows that a leading asymptotic behavior is obtained in this way, but not necessarily that there are not other contributions on the same order, i.e. leading as well. However, these partial contributions to the correlators are distinguished by the fact that they give rise to primitive vertices in the effective action, i.e. vertices that cannot be obtained by gluing lower-order vertices and propagators from the effective action in the scalar sector (see section \ref{s4}). \par 
We discard higher spin operators in the $OPE$, that would carry outside the pure scalar sector, because they would propagate higher-spin particles in the intermediate states. \par
Employing the asymptotic three-point correlator in Eq.(\ref{AS}), a leading asymptotic contribution as $x_1 \rightarrow x_2$ to the four-point correlator in the scalar sector reads:
\bea \label{four0}
\braket{ \mathcal{O}(x_1) \mathcal{O}(x_2)    \mathcal{O}(x_3) \mathcal{O}(x_4)}_{\mathit{conn}} \sim   C(x_1-x_2) C(x_2-x_3) C(x_{3}-x_4) C(x_4-x_2)  \nonumber \\
\sim C(x_1-x_2)  C(x_2-x_3)  C(x_3-x_4)C(x_4-x_1) \nonumber \\
\eea
where in the last line we used $x_1 \sim x_2$. \par
Proceeding by induction, the $r$-point correlator in the scalar sector contains as a leading asymptotic contribution:
\bea \label{r}
\braket{ \mathcal{O}(x_1) \mathcal{O}(x_2)    \cdots \mathcal{O}(x_r) }_{\mathit{conn}} \sim   C(x_1-x_2) \cdots C(x_r-x_1) 
\eea
where again we used $x_1 \sim x_2$ to close the polygon. \par
In fact, in the glueball sector of large-$N$ $QCD$ it is interesting to identify explicitly the Wick contractions corresponding to the Gaussian integration in the free limit, that give rise to this kind of contributions in the $OPE$. \par 
For example we may choose in the scalar glueball sector $\mathcal{O}= Tr(F^2)^k$ for $k>1$, since for $k=1$ $C_3=0$ (see section \ref{s11}). \par
An inspection of the three-point correlator at lowest order shows that $C(x)$ arises at lowest order by Wick contractions
in which the three $2k$-vertices associated to $Tr(F^2)^k$ in perturbation theory form a triangle in which each vertex is linked to a consecutive one by exactly $k$ propagators. \par
This structure extends to the corresponding $r$-polygon in the conformal limit of the $r$-point correlator: The contribution of the $OPE$ to the $r$-point correlator in Eq.(\ref{r}) is matched by the Wick contractions in which each vertex is linked to a consecutive one by exactly $k$ propagators. But the Wick contractions produce all the possible non-equivalent cyclic orderings of the space-time points in Eq.(\ref{r}), equivalent cyclic orderings and inversion of orientation corresponding to the same Wick contractions.  Hence in the algebra of free bosonic fields that occurs at lowest order there are contributions with suitable permutations of the arguments in Eq.(\ref{r}). This plays a role at the end of section \ref{s5}. \par

\section{Asymptotic scalar effective action} \label{s4}

In this section we construct explicitly the asymptotic non-local effective action whose two- and three-point correlators are asymptotic to the result found in the previous sections. \par
From Eq.(\ref{002}) and Eq.(\ref{3}) it follows the asymptotic effective action at lower orders, that reproduces the corresponding correlators integrating on the fields $\Phi_n$ in the $\frac{1}{N}$ expansion, by means of the identification
$\mathcal{O}(x)=\sum_n \Phi_n(x)$:
\bea \label{eff}
\Gamma= &&  \frac{1}{2!} \sum_n \int dq_1 dq_2 \delta(q_1+q_2) m_n^{4-2D} Z_n^{-2}\rho(m^{2}_{n}) \Phi_n(q_1) (q_1^2+m_n^2) \Phi_n(q_2) \nonumber \\
&&+ \frac{C}{3! N^g} \int dq_1 dq_2 dq_3 \delta(q_1+q_2+q_3)        
   \int \sum_{n_1=1}^{\infty} m_{n_1}^2 \frac{m^{-D}_{n_1} Z^{-1}_{n_1} \Phi_{n_1}(q_2)}{p^2+m^{2}_{n_1}  } \nonumber\\
 &&\sum_{n_2=1}^{\infty}   m_{n_2}^2 \frac{m^{-D}_{n_2} Z^{-1}_{n_2}\Phi_{n_2}(q_3)}{(p+q_2)^2+m^{2}_{n_2}  }   
   \sum_{n_3=1}^{\infty}      m_{n_3}^2\frac{m^{-D}_{n_3} Z^{-1}_{n_3}\Phi_{n_3}(q_1)}{(p+q_2+q_3)^2+m^{2}_{n_3}  }dp \nonumber \\
\eea
where $C$ is an $O(1)$ normalization factor, that can be computed in lowest order perturbation theory, the Fourier-transformed field $\Phi_n(q)$ has dimension $D-4$, and the local field $\Phi_n(x)$ dimension $D$. \par
Besides, Fourier transforming Eq.(\ref{r}) and employing the refined Kallen-Lehmann representation for $C(x)$ as for the improved three-point vertex, we get the spectral representation of (some) asymptotic primitive $r$-point vertices in the effective action up to overall normalization:
\bea \label{hl}
 &&\int dq_1 dq_2 \cdots dq_r \delta(q_1+q_2+ \cdots + q_r)        
   \int \sum_{n_1=1}^{\infty} m_{n_1}^2 \frac{m^{-D}_{n_1} Z^{-1}_{n_1} \Phi_{n_1}(q_2)}{p^2+m^{2}_{n_1}  } 
 \sum_{n_2=1}^{\infty}   m_{n_2}^2 \frac{m^{-D}_{n_2} Z^{-1}_{n_2}\Phi_{n_2}(q_3)}{(p+q_2)^2+m^{2}_{n_2}  }   \nonumber \\
&&\cdots   \sum_{n_r=1}^{\infty}      m_{n_r}^2\frac{m^{-D}_{n_r} Z^{-1}_{n_r}\Phi_{n_r}(q_1)}{(p+q_2+ \cdots +q_r)^2+m^{2}_{n_r}  }dp
\eea

\section{Asymptotic $S$-matrix scalar amplitudes } \label{s5}

The asymptotic $S$-matrix three-point amplitude follows from the three-point correlator, or equivalently from the effective action, applying the $LSZ$ reduction formulae, i.e. dividing the on-shell amputated correlator by the product over the external lines of the square root of the 
residues of the propagators, i.e. by $m_n^{D-2} Z_n \rho^{-\frac{1}{2}}(m^{2}_{n})$. The three-point amplitude is therefore:
\bea
S^{(3)}(n_1,n_2,n_3) \sim && \delta(q_1+q_2+q_3)     
   \int   \frac{\rho^{-\frac{1}{2}}(m^{2}_{n_1})}{p^2+m^{2}_{n_1}  } 
   \frac{\rho^{-\frac{1}{2}}(m^{2}_{n_2})} {(p+q_2)^2+m^{2}_{n_2}  }   
    \frac{\rho^{-\frac{1}{2}}(m^{2}_{n_3})}{(p+q_2+q_3)^2+m^{2}_{n_3}  }dp \nonumber \\
 &&   + permutations
\eea
Remarkably, $S^{(3)}$ does not depend on the canonical dimension and on the anomalous dimension of the interpolating field, as it must be. In fact, we show below that all the $S$-matrix amplitudes implied by the effective action in 
Eq.(\ref{eff}) and Eq.(\ref{hl}) do not depend on the naive dimensions and anomalous dimensions.\par
For example one contribution to the four-point amplitude reads, up to normalization:
\bea \label{four}
S^{(4)}(n_1,n_2,n_3,n_4) \sim &&
\delta(q_1+q_2+q_3+q_4)  \sum_n   
   \int   \frac{\rho^{-\frac{1}{2}}(m^{2}_{n_1})}{p^2+m^{2}_{n_1}  } 
   \frac{\rho^{-\frac{1}{2}}(m^{2}_{n_2})} {(p+q_2)^2+m^{2}_{n_2}  }   
    \frac{1}{(p-q_1)^2+m^{2}_{n} }dp \nonumber \\
&&    \frac{\rho^{-1}(m^{2}_{n})}{(q_1+q_2)^2+m^{2}_{n} }  \int   \frac{\rho^{-\frac{1}{2}}(m^{2}_{n_3})}{p^2+m^{2}_{n_3}  } 
   \frac{\rho^{-\frac{1}{2}}(m^{2}_{n_4})} {(p+q_4)^2+m^{2}_{n_4}  }   
    \frac{1}{(p-q_3)^2+m^{2}_{n}  }dp \nonumber \\
 &&   + permutations 
    \eea
Yet for consistency at this order it is necessary to add the contribution of the primitive four-point $S$-matrix amplitude, that up to normalization reads:
\bea 
&& \delta(q_1+q_2 +q_3+q_4)        
  \int  \frac{   \rho_0^{-\frac{1}{2}}(m^2_{n_1})}{p^2+m^{2}_{n_1}  } 
  \frac{ \rho_0^{-\frac{1}{2}}(m^{2}_{n_2})}{(p+q_2)^2+m^{2}_{n_2}  }   \nonumber \\
&&    \frac{ \rho_0^{-\frac{1}{2}}(m^2_{n_3})}{(p+q_2+q_3)^2+m^{2}_{n_3}  }
   \frac{ \rho_0^{-\frac{1}{2}}(m^2_{n_4})}{(p+q_2+q_3+q_4)^2+m^{2}_{n_4}  } dp \nonumber \\
  && +permutations
\nonumber \\
\eea
If we are interested only in the generating functional of the $S$-matrix, but not in the generating functional of correlators, we can rescale the field $\Phi_n$ in the effective action in such a way that acquires canonical dimension, i.e. $1$ in four-dimensional
space-time, and that the kinetic term is canonically normalized. \par
This is equivalent to dividing by the square root of the residues of the propagators in the $LSZ$ formulae. The effective action in canonical form follows:
\bea \label{can}
S= &&  \frac{1}{2!} \sum_n \int dq_1 dq_2 \delta(q_1+q_2) \Phi_n(q_1) (q_1^2+m_n^2) \Phi_n(q_2) \nonumber \\
&&+ \frac{C}{3!N^g } \int dq_1 dq_2 dq_3 \delta(q_1+q_2+q_3)        
   \int \sum_{n_1=1}^{\infty} \frac{   \rho^{-\frac{1}{2}}(m^2_{n_1})\Phi_{n_1}(q_2)}{p^2+m^{2}_{n_1}  } \nonumber\\
 &&\sum_{n_2=1}^{\infty}   \frac{ \rho^{-\frac{1}{2}}(m^{2}_{n_2})\Phi_{n_2}(q_3)}{(p+q_2)^2+m^{2}_{n_2}  }   
   \sum_{n_3=1}^{\infty}  \frac{ \rho^{-\frac{1}{2}}(m^2_{n_3})\Phi_{n_3}(q_1)}{(p+q_2+q_3)^2+m^{2}_{n_3}  }dp +\cdots \nonumber \\
\eea
and analogously for the primitive multi-point vertices, up to normalization, in the generating functional of the $S$-matrix amplitudes:
\bea \label{hlS}
&&\int dq_1 dq_2 \cdots dq_r \delta(q_1+q_2 + \cdots+q_r)        
  \int \sum_{n_1=1}^{\infty} \frac{   \rho_0^{-\frac{1}{2}}(m^2_{n_1})\Phi_{n_1}(q_2)}{p^2+m^{2}_{n_1}  } 
 \sum_{n_2=1}^{\infty}   \frac{ \rho_0^{-\frac{1}{2}}(m^{2}_{n_2})\Phi_{n_2}(q_3)}{(p+q_2)^2+m^{2}_{n_2}  }   \nonumber \\
&&   \cdots \sum_{n_r=1}^{\infty}  \frac{ \rho_0^{-\frac{1}{2}}(m^2_{n_r})\Phi_{n_r}(q_1)}{(p+q_2+\cdots+q_r)^2+m^{2}_{n_r}  }dp  \nonumber \\
\eea
Remarkably, the dependence on the naive and anomalous dimensions has disappeared to all orders in the $\frac{1}{N}$ expansion. This fact opens the way to find an actual solution for the $S$-matrix of large-$N$ $QCD$ (in the scalar sector) provided the spectral information is found by other methods. \par
In fact, employing the conventions displayed in section \ref{s0}, we can write the generating functional of the asymptotic $S$-matrix in the scalar glueball sector of positive charge conjugation in large-$N$ $QCD$ that resums to all orders in the $\frac{1}{N}$ expansion the aforementioned primitive multi-point vertices:
 \bea \label{S1}
S = \frac{1}{2} tr \int   \Phi(-\Delta +  M^2)  \Phi  \, d^4x + \frac{\kappa}{2} N^{2} \log Det_3 (- \Delta +  M^2 - \frac{c'} { N} \rho_0^{-\frac{1}{2}} \Phi)
\eea
The argument is as follows. The loop expansion of the logarithm of the square root of the functional determinant in Eq.(\ref{S1}) carries the factor of $\frac{1}{2r}$ for the primitive $r$-point vertex, with the factor of $\frac{1}{r}$ coming from the logarithm and the factor of $\frac{1}{2}$ from the square root. Thus in the effective action in Eq.(\ref{S1}) the $r$-point vertex
is divided by the
number $r$ of cyclic permutations of the $r$ vertices of the $r$-polygon, and by the number $2$ of orientations of the boundary of the polygon, but it is invariant precisely for these transformations. Besides, all the permutations are generated by Wick contractions for the $r$-point correlator in terms of the field $\Phi$. Thus the correlator that arises from the effective action (after scaling back the fields to the appropriate non-canonical normalization) contains a sum over all permutations modulo cyclic permutations and orientation inversion. \par
But this combinatorics, that is implied by the normalization of the $r$-vertices in the effective action, matches the combinatorics of the Wick contractions arising by the algebra
of free bosonic fields in the conformal limit described in the section \ref{s3}, with the overall factor of $\kappa$ in front of the logarithm of the functional determinant defined by the relation:
\bea
G^{(2)}(x)=\kappa \, C^2(x) 
\eea
asymptotically, and with the normalization of $C(x)$ defined by the $OPE$. \par
Indeed, by Eq.(\ref{OPE}) and Eq.(\ref{r}) each contribution to the $r$-correlator is given by $r$ factors of $C(x_i-x_j)$ for certain permutations of the space-time coordinates and an overall factor of $\kappa$. Thus the question is which permutations of the cyclically ordered space-time points occur in these factors of $C(x_i-x_j)$ due to the $OPE$. But we have just noticed that there is a one-to-one correspondence between the Wick contractions in the free limit in perturbation theory that must match the $OPE$, that generate all the permutations of the space-time points modulo cyclic permutations and inversion, and the Wick contractions arising from the $\frac{1}{N}$ expansion of the functional determinant in the effective action. \par
To complete the argument about the effective action we observe that in general the normalization of $C(x)$, as it arises in the $OPE$, does not correspond to a canonical normalization of the kinetic term in the effective action. This is taken into account by the factor of $c'$ in Eq.(\ref{S1}). \par
The effective action in the scalar sector is manifestly super-renormalizable by power counting because the density $\rho_0^{-\frac{1}{2}}$ has dimension of $\Lambda_{QCD}$. Besides, all the primitive scalar vertices in the $S$-matrix are finite by power counting.

\section{Spin-$1$ and higher-spin correlators} \label{s9}

 Contrary to the scalar case, for which the structure of the three-point correlators is uniquely determined up to normalization in the conformal limit,
 for higher spins the number of inequivalent conformal structures increases with spin according to the classification in section 3 of \cite{Stanev}. \par
 In massless $QCD$ and $\mathcal{N}=1$ $SUSY$ $YM$ the correlators in the conformal limit
 are all realized as correlators of composite operators in a free massless theory: Maxwell and Weyl fermion. \par 
Therefore, from a practical point of view and for our purposes, it is necessary to proceed case by case in order to compute the conformal limit, taking advantage that both the general structures \cite{Stanev} and some interesting particular cases
 \cite{Stanev,Todorov} have been worked out in detail already. \par
 In particular in \cite{Todorov} it has been found that  several conformal correlators, that are physically very relevant from our point of view for $QCD$, simplify greatly in spinor, as opposed to vector, notation. Therefore, when it is convenient, we adopt the spinor notation in the following sections. 
\subsection{Spin-$1$ effective action}

The simplest three-point correlators next to the scalar ones involve vector, axial and chiral currents, built by means of fermion bilinears in $QCD$ (and in $SUSY$ $YM$). From the point of view of the conformal limit this is the Weyl theory. \par
Since we have already worked out the fundamental principles of the asymptotically-free bootstrap in the scalar case, for spin-$1$ currents it is convenient to display directly the final answer for the generating functional of the correlators, and to check only a posteriori that it reproduces the correct correlators and $OPE$, the basic additional difficulty with respect to the scalar case being the combinatorics of the extra indices. \par
To set the notation we introduce flavor chiral $R=+,-$, vector $R=V$, and axial $R=A$ currents, $j^{a}_{R \alpha \dot \beta}$ with $a$ valued in the Lie algebra of $U(N_f)$ defined in section \ref{s7}, with $a$ a flavor index in the Lie algebra of
$SU(N_f)$ for non-singlets, and $a=0$ for flavor singlets (i.e. the $U(1)$ part of $U(N_f)$). Besides, we suitably normalize the currents, in order to get a canonically defined large-$N$ limit.\par
The generating functional of spin-$1$ correlators at lower orders reads in spinor notation:
\bea \label{effspin1}
&&\Gamma_{1R} = \frac{1}{2} \sum_n \int dq_1 dq_2 \delta(q_1+q_2) m_{Rn}^{-2} Z_{Rn}^{-2}  \rho_{1R}(m^2_{Rn})  \delta_{ab} \nonumber \\
&&\Phi^{a \alpha_1 \dot \beta_1}_{Rn }(q_1) \big(\epsilon_{\alpha_1 \alpha_2} \epsilon_{\dot \beta_1 \dot \beta_2}(q_1^2+ m_{Rn}^2) - q_{1 \alpha_1 \dot \beta_1} q_{1 \alpha_2 \dot \beta_2}  \big) \Phi^{b \alpha_2 \dot \beta_2}_{Rn}(q_2) \nonumber \\
&&+ \frac{C}{3 N^g} \int dq_1 dq_2 dq_3 \delta(q_1+q_2+q_3)   Tr^{(R)}(a,b,c)  
 \int \sum_{n_1=1}^{\infty}  \frac{p_{\alpha_1 \dot \beta_2} m^{-2}_{Rn_1} z^{-1}_{Rn_1} \Phi^{a \alpha_1 \dot \beta_1}_{Rn_1}(q_2)}{p^2+m^{2}_{Rn _1}  } \nonumber \\
&& \sum_{n_2=1}^{\infty}  \frac{(p+q_2)_{\alpha_2 \dot \beta_3} m^{-2}_{Rn_2} z^{-1}_{Rn_2} \Phi^{b \alpha_2 \dot \beta_2}_{Rn_2}(q_3)}{(p+q_2)^2+m^{2}_{Rn _2}  }      
 \sum_{n_3=1}^{\infty}  \frac{(p+q_2+q_3)_{\alpha_3 \dot \beta_1} m^{-2}_{Rn_3} z^{-1}_{Rn_3} \Phi^{c \alpha_3 \dot \beta_3}_{Rn_3}(q_1)}{(p+q_2+q_3)^2+m^{2}_{Rn _3}  } dp + \cdots \nonumber \\
\eea
where the sums involve only states of type $R$, and:
\bea \label{cd}
&& \lim_{n \rightarrow \infty} Z_{Rn} =1 \nonumber \\
  && \sum_n z_{Rn} m_{Rn}^{-2}  \rho^{-1}_{1}(m^2_{Rn}) \sim 1 \nonumber \\
  &&Tr^{(+)}(a,b,c)= Tr(T^aT^bT^c)\nonumber \\
  &&Tr^{(-)}(a,b,c)= - Tr(T^cT^bT^a)\nonumber \\
  && Tr^{(V)}(a,b,c)=Tr([T^a,T^b]T^c) \sim f^{abc} \nonumber \\
  && Tr^{(A)}(a,b,c)=Tr(\{T^a,T^b\}T^c) \sim d^{abc} \nonumber \\
  \eea
with $C \sim O(1)$ computable in lowest-order perturbation theory, and $g=\frac{1}{2}$ for mesons, $g=1$ for gluinoballs (for which of course there is no flavor).  \par
Some observations are in order. \par
The correlators are computed by the identification $j^{a\alpha_1 \dot \beta_1}_{R }(x)=\sum_n \Phi^{a \alpha_1 \dot \beta_1}_{Rn}$. \par
In 't Hooft large-$N$ limit all the $R$ currents are conserved at the leading $\frac{1}{N}$ order, therefore $\gamma_0$ vanishes, and $\lim_{n \rightarrow \infty} Z_{Rn}=1$ in the kinetic term of the effective action. \par
It follows from the asymptotic theorem that the two-point spin-$1$ correlators:
\bea \label{2spin1}
&& <j^{a}_{R \alpha_1 \dot \beta_1}(x) j^{b}_{R \alpha_2 \dot \beta_2}(0)>  \nonumber \\
&&\sim \delta^{ab} \sum_n \frac{1}{(2 \pi)^4} \int 
  \big(\epsilon_{\alpha_1 \alpha_2} \epsilon_{\dot \beta_1 \dot \beta_2} + \frac{p_{ \alpha_1 \dot \beta_1} p_{ \alpha_2 \dot \beta_2}}{m_{Rn}^2}) 
   \frac{  m_{Rn}^{2} Z_{Rn}^{2}   \rho_{1R}^{-1}(m^2_{Rn})}{p^2+m^{2}_{Rn} } e^{ip \cdot x} dp \nonumber \\
   &&= \delta^{ab} \frac{1}{(2 \pi)^4}  \int p^2
  \big(\epsilon_{\alpha_1 \alpha_2} \epsilon_{\dot \beta_1 \dot \beta_2} - \frac{p_{ \alpha_1 \dot \beta_1} p_{ \alpha_2 \dot \beta_2}}{p^2}) 
  \sum_n \frac{  Z_{Rn}^{2}   \rho_{1R}^{-1}(m^2_{Rn})}{p^2+m^{2}_{Rn } } e^{ip \cdot x} dp+ \cdots \nonumber \\
\eea
are transverse also off-shell, up to divergent contact terms supported on $x=0$ (see section \ref{s11}). \par
By the asymptotic theorem in section \ref{s11}, Eq.(\ref{2spin1}) is asymptotic to the $RG$-improved perturbative result:
\bea
&& <j^{a}_{R \alpha_1 \dot \beta_1}(x) j^{b}_{R \alpha_2 \dot \beta_2}(0)>  \sim \delta^{ab} \frac{x_{\alpha_1 \dot \beta_2} x_{\alpha_2 \dot \beta_1}}{x^8}
\eea
that coincides with the conformal limit \cite{Todorov}, since the currents are conserved. \par
In fact, the divergence of the flavor singlet-axial current in $QCD$ mixes with the operator $Tr(F_{AB} \,^{*}\!F_{AB})$ at next-to-leading $\frac{1}{N}$ order. The effective action in Eq.(\ref{effspin1}) does not take into account this mixing, because it is on the order of $g^2(\mu)$ in perturbation theory and the operator $Tr(F_{AB} \,^{*}\!F_{AB})$ has anomalous dimension $2 \beta_0$. Therefore, the mixing is subleading from the point of view of the $UV$ asymptotics. \par
The convergence condition $\sum_n z_{Rn} m_{Rn}^{-2}  \rho^{-1}_{1}(m^2_{Rn}) \sim 1$ in the three-point vertex is needed to imply the correct asymptotics of the coefficient function of the current $j_R$ itself in the  $OPE$ of two currents, and it occurs as a special feature for conserved currents with $D=3$ (see section \ref{s11}):
\bea \label{OPESPIN1}
\frac{1}{(2 \pi)^4} \int \sum_{n=1}^{\infty}  \frac{p_{\alpha \dot \beta} m^{-2}_{Rn} z_{Rn} \rho^{-1}_{1R}(m^2_{Rn})   }{p^2+m^{2}_{Rn}  }  e^{ip \cdot x} dp \sim   \frac{x_{\alpha \dot \beta}} {x^4} 
\eea
The three-point vertex in Eq.(\ref{effspin1}) is divergent by power counting because of the loop integration $dp$. Thus regularization is required, that leads eventually to the standard triangle anomaly, but to finite vertices.\par
From the effective action it follows the asymptotic spectral representation of the three-point correlator:
\bea \label{v}
&& <j^{a}_{R \alpha_1 \dot \beta_1}(q_1) j^{b}_{R \alpha_2 \dot \beta_2}(q_2) j^{c}_{R \alpha_3 \dot \beta_3}(q_3)>  \nonumber \\
&& \sim \frac{1}{ N^g} \delta(q_1+q_2+q_3)   Tr^{(R)}(a,b,c)  
 \int \sum_{n_1=1}^{\infty}  \frac{p_{\alpha_1 \dot \beta_2} m^{-2}_{Rn_1} z_{Rn_1}    \rho^{-1}_{1R}(m^2_{Rn_1})}{p^2+m^{2}_{Rn _1}  } \frac{m_{Rn_1}^2}{q_2^2+m_{Rn_1}^2} \nonumber \\
&& \sum_{n_2=1}^{\infty}  \frac{(p+q_2)_{\alpha_2 \dot \beta_3} m^{-2}_{Rn_2} z_{Rn_2}    \rho^{-1}_{1R}(m^2_{Rn_2}) }{(p+q_2)^2+m^{2}_{Rn _2}  }      \frac{m_{Rn_2}^2}{q_3^2+m_{Rn_2}^2} \nonumber \\
&& \sum_{n_3=1}^{\infty}  \frac{(p+q_2+q_3)_{\alpha_3 \dot \beta_1} m^{-2}_{Rn_3} z_{Rn_3}\rho^{-1}_{1R}(m^2_{Rn_3})}{(p+q_2+q_3)^2+m^{2}_{Rn _3}   } \frac{m_{Rn_3}^2}{q_1^2+m_{Rn_3}^2} dp \nonumber \\
&&+ opposite \, orientation
\eea
where, to keep the notation simple, we have omitted the transverse projectors in the external lines. For vector currents the correlator is automatically asymptotically transverse after suitable regularization. For chiral and axial currents the longitudinal part of the asymptotic correlator, being asymptotically proportional to the divergence of the triangle anomaly, is a polynomial in momentum, i.e. a contact term. \par
The last term in Eq.(\ref{v}) is due to summing on non-cyclic permutations of the vertex (the vertex is cyclically symmetric) in the effective action.\par
The three-point correlators must be asymptotic, as we actually show below, to the $RG$-improved perturbative result, i.e. to the triangle graphs in perturbation theory.  \par 
Asymptotically, as in the scalar case we may skip the factors asymptotic to $1$ associated to the external lines, to get:
\bea \label{am1}
&& <j^{a}_{R \alpha_1 \dot \beta_1 }(q_1) j^{b}_{R\alpha_2 \dot \beta_2}(q_2) j^{c}_{R\alpha_3 \dot \beta_3 }(q_3)>  \nonumber \\
&& \sim \frac{1}{ N^g}  \delta(q_1+q_2+q_3)   Tr^{(R)}(a,b,c)  
 \int \sum_{n_1=1}^{\infty}  \frac{p_{\alpha_1 \dot \beta_2} m^{-2}_{Rn_1} z_{Rn_1}  \rho^{-1}_{1R}(m^2_{Rn_1})  }{p^2+m^{2}_{Rn _1}   }  \nonumber \\
&& \sum_{n_2=1}^{\infty}  \frac{(p+q_2)_{\alpha_2 \dot \beta_3} m^{-2}_{Rn_2} z_{Rn_2}   \rho^{-1}_{1R}(m^2_{Rn_2}) }{(p+q_2)^2+m^{2}_{Rn _2}  }      
 \sum_{n_3=1}^{\infty}  \frac{(p+q_2+q_3)_{\alpha_3 \dot \beta_1} m^{-2}_{Rn_3} z_{Rn_3}   \rho^{-1}_{1R}(m^2_{Rn_3})  }{(p+q_2+q_3)^2+m^{2}_{Rn _3}  } dp \nonumber \\
&&+ opposite \, orientation
\eea
Employing Eq.(\ref{OPESPIN1}) (see section \ref{s11}) and Fourier transforming the convolution in Eq.(\ref{am1}) into a product in the coordinate representation as in the scalar case, we get the $RG$-improved asymptotic results in perturbation theory as follows. \par
The three-point correlator of the chiral current $R=+$ reads asymptotically:
\bea
&& <j^{a\alpha_1 \dot \beta_1}_{+ }(x_1) j^{b\alpha_2 \dot \beta_2}_{+ }(x_2) j^{c\alpha_3 \dot \beta_3}_{+ }(x_3)>  \nonumber \\
&&\sim Tr(T^aT^bT^c)  \frac{x^{\alpha_1 \dot \beta_2}_{12}} {x_{12}^4}   \frac{x^{\alpha_2 \dot \beta_3}_{23}}{x_{23}^4}\frac{x^{\alpha_3 \dot \beta_1}_{31}}{x_{31}^4}
+Tr(T^aT^cT^b)  \frac{x^{\alpha_1 \dot \beta_3}_{13}} {x_{13}^4}   \frac{x^{\alpha_3 \dot \beta_2}_{32}}{x_{32}^4}\frac{x^{\alpha_2 \dot \beta_1}_{21}}{x_{21}^4}
\eea
that coincides with the conformal limit \cite{Todorov} because $\gamma_0=0$ for all the $R$ currents at leading $\frac{1}{N}$ order. \par 
In perturbation theory the two terms correspond to the two orientations of the triangle that arises performing fermion integration
in the free Weyl theory. \par
For $R=-$, employing the computational technology in \cite{Weyl} for chiral fermions in spinor notation, we get a result that differs by an overall sign and by the transposition of the flavor matrices:
\bea
&& <j^{a\alpha_1 \dot \beta_1}_{- }(x_1) j^{b\alpha_2 \dot \beta_2}_{- }(x_2) j^{c\alpha_3 \dot \beta_3}_{- }(x_3)>  \nonumber \\
&& \sim - Tr(T^c T^b  T^a)  \frac{x^{\alpha_1 \dot \beta_2}_{12}} {x_{12}^4}   \frac{x^{\alpha_2 \dot \beta_3}_{23}}{x_{23}^4}\frac{x^{\alpha_3 \dot \beta_1}_{31}}{x_{31}^4}
- Tr(T^b T^c T^a)  \frac{x^{\alpha_1 \dot \beta_3}_{13}} {x_{13}^4}   \frac{x^{\alpha_3 \dot \beta_2}_{32}}{x_{32}^4}\frac{x^{\alpha_2 \dot \beta_1}_{21}}{x_{21}^4} \nonumber \\
\eea
Given these basic correlators, the three-point correlators of vector and axial currents are obtained taking the appropriate linear combinations. \par
We get asymptotically for the $VVV$ and $VAA$ correlators \cite{Todorov}:
\bea
&& <j^{a\alpha_1 \dot \beta_1}_{V}(x_1) j^{b\alpha_2 \dot \beta_2}_{V }(x_2) j^{c\alpha_3 \dot \beta_3}_{V }(x_3)>  \nonumber \\
&& \sim <j^{a\alpha_1 \dot \beta_1}_{V}(x_1) j^{b\alpha_2 \dot \beta_2}_{A }(x_2) j^{c\alpha_3 \dot \beta_3}_{A }(x_3)>  \nonumber \\
&&\sim f^{abc}  (\frac{x^{\alpha_1 \dot \beta_2}_{12}} {x_{12}^4}   \frac{x^{\alpha_2 \dot \beta_3}_{23}}{x_{23}^4}\frac{x^{\alpha_3 \dot \beta_1}_{31}}{x_{31}^4}
-  \frac{x^{\alpha_1 \dot \beta_3}_{13}} {x_{13}^4}   \frac{x^{\alpha_3 \dot \beta_2}_{32}}{x_{32}^4}\frac{x^{\alpha_2 \dot \beta_1}_{21}}{x_{21}^4}) 
\eea
and for the $AAA$ and $AVV$ correlators \cite{Todorov}:
\bea
&& <j^{a\alpha_1 \dot \beta_1}_{A}(x_1) j^{b\alpha_2 \dot \beta_2}_{A }(x_2) j^{c\alpha_3 \dot \beta_3}_{A }(x_3)>  \nonumber \\
&& \sim <j^{a\alpha_1 \dot \beta_1}_{A}(x_1) j^{b\alpha_2 \dot \beta_2}_{V }(x_2) j^{c\alpha_3 \dot \beta_3}_{V }(x_3)>  \nonumber \\
&&\sim d^{abc}  (\frac{x^{\alpha_1 \dot \beta_2}_{12}} {x_{12}^4}   \frac{x^{\alpha_2 \dot \beta_3}_{23}}{x_{23}^4}\frac{x^{\alpha_3 \dot \beta_1}_{31}}{x_{31}^4}
+ \frac{x^{\alpha_1 \dot \beta_3}_{13}} {x_{13}^4}   \frac{x^{\alpha_3 \dot \beta_2}_{32}}{x_{32}^4}\frac{x^{\alpha_2 \dot \beta_1}_{21}}{x_{21}^4})
\eea
Moreover, setting the kinetic term in canonical form, the generating functional of the asymptotic $S$-matrix at lower orders is:
\bea \label{Sspin1}
S_{1R}= &&  \frac{1}{2} \sum_n \int dq_1 dq_2 \delta(q_1+q_2)  \delta_{ab}
 \Phi^{a \alpha_1 \dot \beta_1}_{Rn }(q_1)  \big(\epsilon_{\alpha_1 \alpha_2} \epsilon_{\dot \beta_1 \dot \beta_2}(q_1^2+ m_{Rn}^2) - q_{1 \alpha_1 \dot \beta_1} q_{1 \alpha_2 \dot \beta_2}  \big) \Phi^{b \alpha_2 \dot \beta_2}_{Rn}(q_2) \nonumber \\
&&+ \frac{C}{3 N^g} \int dq_1 dq_2 dq_3 \delta(q_1+q_2+q_3)   Tr^{(R)}(a,b,c)    \nonumber \\
&&   \int \sum_{n_1=1}^{\infty}  \frac{p_{\alpha_1 \dot \beta_2} Z_{Rn_1} z_{Rn_1}^{-1} m^{-1}_{Rn_1}  \rho^{-\frac{1}{2}}_{1R}(m^2_{Rn_1}) \Phi^{a \alpha_1 \dot \beta_1}_{Rn_1}(q_2)}{p^2+m^{2}_{Rn _1}  } \nonumber\\
 &&  \sum_{n_2=1}^{\infty}  \frac{(p+q_2)_{\alpha_2 \dot \beta_3} Z_{Rn_2} z_{Rn_2}^{-1} m^{-1}_{Rn_2} \rho^{-\frac{1}{2}}_{1R}(m^2_{Rn_2}) \Phi^{b \alpha_2 \dot \beta_2}_{Rn_2}(q_3)}{(p+q_2)^2+m^{2}_{Rn _2}  }     \nonumber \\   
&& \sum_{n_3=1}^{\infty}  \frac{(p+q_2+q_3)_{\alpha_3 \dot \beta_1} Z_{Rn_3} z_{Rn_3}^{-1}m^{-1}_{Rn_3} \rho^{-\frac{1}{2}}_{1R}(m^2_{Rn_3}) \Phi^{c \alpha_3 \dot \beta_3}_{Rn_3}(q_1)}{(p+q_2+q_3)^2+m^{2}_{Rn _3}  } dp + \cdots \nonumber \\
\eea
The factors of $ p_{\alpha_1 \dot \beta_2} m^{-1}_{Rn_1}$ spoil the super-renormalizability by power counting of the $S$-matrix, because now the effective coupling is dimensionless $\rho^{-\frac{1}{2}}_{1R}  m^{-1}_{Rn} $.
Thus, by power counting only, in the spin-$1$ sector (for fermion bilinears) the large-$N$ theory is renormalizable but not super-renormalizable. For spin-$1$ there are ambiguities similar to the scalar case in the choice of contact terms in the coefficient of the $OPE$. 

\subsection{Vector dominance, pion form factor, light by light scattering}

For $R=V$ the convergence condition in the three-point vertex in the spin-$1$ effective action $\sum_n z_{Rn} m_{Rn}^{-2}  \rho^{-1}_{1}(m^2_{Rn}) \sim 1 $ is related to the famous vector dominance in the pion form factor, i.e. the fact that, due to this condition in the correlator in Eq.(\ref{am1}), the higher states are quite suppressed with respect to the lowest state, e.g. the rho meson in the charged pion form factor.\par
 Indeed, both the charged and neutral (asymptotic) pion form factors can be extracted from the (asymptotic) spectral representation of the $VAA$ and $VVA$ correlators, employing the axial currents to create on shell 
(massless) pions from the vacuum. This will be worked out in more detail elsewhere. \par
It is convenient to explore the consequences of the aforementioned convergence condition supposing first that the series converges absolutely. This is equivalent to the convergence of the integral $\int  z_R(m_R) m_R^{-2} dm_R^2 \sim 1 $ for large masses, that requires
that $z_R(m_R)$ vanishes for large $m_R$ at least as $z_R(m_R) \sim \frac{1}{\log^{1+\epsilon}(m_R^2)}$ with $\epsilon > 0$. However, if the series does not converge absolutely, it suffices \footnote{Strictly speaking this is certainly sufficient only for $\rho_{1}(m^2_{Rn}) \sim \Lambda_{QCD}^{-2}$, i.e. for a spectral distribution asymptotically linear in the masses squared.} that 
$z_{Rn} \sim (-1)^n$, i.e. that the series is alternating as pointed out firstly by
Shifman \cite{Svector}, otherwise for $R^{(2)}_{Rn}=z_{Rn} \sim 1$, the series in Eq.(\ref{R}) would diverge logarithmically contrary to the asymptotic solution. \par
We stress that in our framework (asymptotic) vector dominance is a consequence of the existence of the asymptotic conformal
limit for the currents and of the Kallen-Lehmann representation in the $OPE$ of the currents (see section \ref{s11}).\par
\par
Finally, the correlator of four vector currents follows from the three-point vertex and the coefficients of the $OPE$. This correlator is relevant for light by light scattering, but this application will be worked out elsewhere. A contribution to the correlator of four currents that is factorized on spin-$1$ vector states \footnote{Of course, there are also other contributions factorized on states of different spin, that can be obtained by the same methods employing the $OPE$.} arises from the primitive four-point vertex, up to perhaps symmetrization:
\bea \label{light}
&& < j^{a}_{V\alpha_1 \dot \beta_1 }(q_1) j^{b}_{V\alpha_2 \dot \beta_2}(q_2) j^{c}_{V\alpha_3 \dot \beta_3}(q_3) j^{d}_{V\alpha_4 \dot \beta_4 }(q_4) >  \sim  \nonumber \\
&& \frac{1}{ N}  \delta(q_1+q_2+q_3+q_4)   \int \Big ( f^{abc'}
 \sum_{n_1=1}^{\infty}  \frac{p_{\alpha_1 \dot \beta_2} m^{-2}_{Vn_1} z_{Vn_1}    \rho^{-1}_{1V}(m^2_{Vn_1})}{p^2+m^{2}_{Vn _1}  } \frac{m_{Vn_1}^2}{q_2^2+m_{Vn_1}^2} \nonumber \\
&& \sum_{n_2=1}^{\infty}  \frac{(p+q_2)_{\alpha_2 \dot \beta'} m^{-2}_{Vn_2} z_{Vn_2}    \rho^{-1}_{1V}(m^2_{Vn_2}) }{(p+q_2)^2+m^{2}_{Vn _2}  }      \frac{m_{Vn_2}^2}{q_3^2+m_{Vn_2}^2} \nonumber \\
&& \sum_{n_3=1}^{\infty}  \frac{(p+q_2+q_3)_{\alpha' \dot \beta_1} m^{-2}_{Vn_3} z_{Vn_3}\rho^{-1}_{1V}(m^2_{Vn_3})}{(p+q_2+q_3)^2+m^{2}_{Vn _3}   } \frac{m_{Vn_3}^2}{q_4^2+m_{Vn_3}^2}  + opposite \, orientation \Big)  \nonumber \\
&&  f^{c'cd} \big( (p+q_2+q_3+q_4)_{\alpha_3 \dot \beta_4} \delta_{\alpha_4}^{ \alpha'}\delta_{\dot \beta_3}^{ \dot \beta'}
+ (p+q_2+q_3+q_4)_{\alpha_4 \dot \beta_3} \delta_{\alpha_3}^{ \alpha'} \delta_{\dot \beta_4}^{ \dot \beta'}) \nonumber \\
&& \sum_{n_4=1}^{\infty}  \frac{ m^{-2}_{Vn_4} z_{Vn_4}  \rho^{-1}_{1}(m^2_{Vn_4})}{(p+q_2+q_3+q_4)^2+m^{2}_{Vn_4 }  }\frac{m_{Vn_4}^2}{q_1^2+m_{Vn_4}^2} 
 dp \nonumber \\
\eea
where in the external lines, to keep the notation simple, we have skipped the transverse projectors.
This four-point correlator is obtained combining the three-point correlator with the $OPE$ (see section \ref{s11}), as in Eq.(\ref{four0}) in the scalar case. \par
At the same order there is another contribution, that arises gluing with a propagator two primitive three-point vertices, analog to Eq.(\ref{four}) in the scalar case. It can be read immediately from the effective action in Eq.(\ref{effspin1}).

\section{Spectrum generating algebra: $QCD$ glueballs and mesons, and $\mathcal{N}=1$ $SUSY$ $YM$ gluinoballs} \label{s7}

In massless $QCD$ scalar and pseudoscalar  three-point correlators constructed by quark bilinears $\bar qT^a q$ and  $\bar q \gamma_5T^a q$ vanish at lowest order of perturbation theory, because the corresponding fermion loop contains an odd number of $\gamma_{A}$ matrices arising from the (massless) quark propagators. The same holds for operators obtained from them by multiple insertions of covariant derivatives, that form a part of the spectrum generating algebra for mesons \cite{Salgebra}. Thus the asymptotic solution, at least in the form worked out in this paper, does not apply to them. \par
Also the three-point correlator of $TrF^2$ vanishes in the conformal limit, as the $OPE$ coefficient of $TrF^2$ starts at order of $g^2$ up to contact terms \cite{chet}. But in general the three-point correlator does not vanish for $Tr(F^2)^k$
with $k >1$. Thus the asymptotically-free bootstrap applies. \par
Three-point correlators of quark bilinears involving vector $\bar qT^a \gamma_A q$, axial $\bar qT^a \gamma_A \gamma_5 q$, and chiral currents $\bar qT^a \gamma_A(1+ \gamma_5) q$, $\bar qT^a \gamma_A(1- \gamma_5) q$, do not vanish in the conformal limit, because the corresponding fermion loop in lowest-order perturbation theory contains an even number of gamma matrices arising by the combination of the gamma matrices in the operators and the (massless) quark propagators. \par
The same holds for operators obtained from them by multiple insertions of covariant derivatives, that belong to a part of the spectrum generating algebra for mesons \cite{Salgebra}, in such a way that the asymptotic solution applies.\par
Moreover, there are scalar fermion bilinears that contain both $\gamma_A$ and $F^{AB}$ \cite{Salgebra}, such as $\bar q \gamma_A F^{AE} F_{E}^B D_Bq$, whose three-point correlators do not vanish in general
in the conformal limit. Their chiral version occurs in Ferretti-Heise-Zarembo integrable sector of large-$N$ $QCD$ \cite{Fe}. \par
Entirely analogous statements hold for chiral gluinoball operators built by fermion bilinears in $\mathcal{N}=1$ $SUSY$ $YM$.

\section{Asymptotic theorem for the coefficients of the $OPE$} \label{s11}

In this section we prove some refinements of the asymptotic theorem \cite{MBN} and its further extension to the coefficient functions of the $OPE$ in the scalar case. In addition, we report a detailed investigation in the spin-$1$ case, because of certain peculiarities with respect to the scalar case. The asymptotic theorem may extend to operator mixing and to the coefficient functions in the $OPE$ of higher-spin operators as well, but this requires a detailed treatment that will appear elsewhere. \par
We assume that the gauge theory that we deal with is asymptotically-free, with no physical mass scale in perturbation theory. As a consequence all the correlators of the theory are conformal in the free limit $g(\mu) \rightarrow 0$, because the beta function vanishes in this limit,
although some of them may vanish in this limit. \par
It is convenient to start at first with the scalar coefficient functions $C_{2D-D_1}$ in the $OPE$ of the scalar operator $\mathcal{O}_D$ of naive dimension $D$ in the coordinate representation:
\bea \label{ope0}
 G^{(2)}(x)=\langle \mathcal{O}_D(x)\mathcal{O}_D(0) \rangle_{conn}= C_{2D}(x)+\sum_{D_1 > 0} C_{2D-D_1}(x) <\mathcal{O}_{D_1}(0)>
\eea
We assume that the scalar condensates do not vanish, that is believed to hold generically in large-$N$ $QCD$. In $\mathcal{N}=1$ $SUSY$ $YM$ by an argument in \cite{Shifman1} the non-vanishing of the condensates requires scalar non-purely gluonic operators. \par
Tautologically, in the coordinate representation \cite{chet2} no contact term, i.e. distributions supported at $x=0$, arises for $x\neq 0$.
Hence the operators $\mathcal{O}_{D}, \mathcal{O}_{D_1}$ are multiplicatively renormalizable in the coordinate representation and therefore the coefficient functions $C_{2D}, C_{2D-D_1}$ in the $OPE$ satisfy \cite{chet2} the Callan-Symanzik equations:
\begin{equation}\label{a}
\left(x_{M}\frac{\partial}{\partial x_{M}}+\beta(g)\frac{\partial}{\partial g}+2(D+\gamma_{\mathcal{O}_D}(g))\right)C_{2D}(x)=0
\end{equation}
and:
\begin{equation}\label{boc}
\left(x_{M}\frac{\partial}{\partial x_{M}}    +\beta(g)\frac{\partial}{\partial g}+(2D-D_1+2\gamma_{\mathcal{O}_D}(g)- \gamma_{\mathcal{O}_{D_1}}(g)  )\right)C_{2D-D_1}(x)=0
\end{equation}
The general solutions are:
\begin{equation}\label{00}
C_{2D}(x)= \frac{1}{x^{2D}}\,\mathcal{G}_{2D}(g(x))\, Z^2_{\mathcal{O}_D}(x \mu,g(\mu))
\end{equation}
and:
\begin{equation} \label{01}
C_{2D-D_1}(x)= \frac{1}{x^{2D-D_1}}\,\mathcal{G}_{2D-D_1}(g(x))\, Z^2_{\mathcal{O}_D}(x \mu,g(\mu)) Z^{-1}_{\mathcal{O}_{D_1}}(x \mu,g(\mu))
\end{equation}
The $RG$-invariant functions $\mathcal{G}_{2D},\mathcal{G}_{2D-D_1}$ of the running coupling $g(x)$ only can be computed order by order in perturbation theory. \par
If the perturbative two-point correlator is non-vanishing in the conformal limit in perturbation theory, as it happens in a unitary theory, i.e. $C_2 \neq 0$ (see section \ref{s2}), the perturbative estimate for $\mathcal{G}_{2D}$ is \cite{MBN}:
\begin{equation}
\mathcal{G}_{2D}(g(x)) \sim 1 + O(g^2(x))
\end{equation}
otherwise $C_2$ would vanish. \par
In the same fashion, if the three-point correlator:
\bea
 \langle \mathcal{O}_D(x)\mathcal{O}_D(0) \mathcal{O}_{D_1}(y)\rangle_{conn}
 \eea
is non-vanishing in the conformal limit, then
\begin{equation}
\mathcal{G}_{2D-D_1}(g(x)) \sim 1 + O(g^2(x))
\end{equation}
Otherwise, $\mathcal{G}_{2D-D_1}(g(x)) \sim O(g^{2l}(x))$ for some positive integer $l$, that can be determined by a direct perturbative computation. \par
Under the previous assumptions, exploiting the asymptotic behavior of the general solution in Eq.(\ref{00}) and Eq.({\ref{01}) \cite{MBN}, we get the actual small-distance asymptotic behavior of the scalar coefficient functions in the scalar $OPE$:
\begin{align}
&C_{2D}(x)\sim \frac{1}{x^{2D}} \,
(\frac{g(x)}{g(\mu)})^{\frac{2\gamma_0(\mathcal{O}_D)  }{\beta_0}}
\end{align}
and:
\begin{align}\label{acf}
&C_{2D-D_1}(x)\sim g(x)^{2l} \frac{1}{x^{2D-D_1}} \,
(\frac{g(x)}{g(\mu)})^{\frac{2\gamma_0(\mathcal{O}_D)  - \gamma_0(\mathcal{O}_{D_1}) }{\beta_0}}
\end{align}
with $l=0$ if the three-point correlator does not vanish in the conformal limit. \par
Now we assume that the theory confines in the large-$N$ 't Hooft limit.
As a consequence the two-point scalar correlator satisfies the Kallen-Lehmann representation:
\bea \label{KL}
G^{(2)}(x)&& = \sum_{n=1}^{\infty} \frac{1}{(2 \pi)^4} \int   \frac{|< 0|\mathcal{O}_D(0)|n>|^2}{p^2+m^{2}_n  } \,e^{ ip\cdot x}d^4p    \nonumber \\
&& = \sum_{n=1}^{\infty}  \frac{1}{(2 \pi)^4} \int  \frac{\mathcal{R}_{n}}{p^2+m^{ 2}_{n}} \,e^{ ip\cdot x}d^4p   
\eea
in which only scalar single-particle states occur, because by assumption in a large-$N$ limit of 't Hooft type the interaction and the particle widths are suppressed at the leading large-$N$ order. \par
Because of confinement \footnote{This is our definition of confinement in the large-$N$ 't Hooft limit (see section \ref{s0}).} we assume that the (scalar) spectrum is a discrete diverging sequence $ \{ m_n \}$ at the leading large-$N$ order. At the same time we assume that the spectrum $ \{ m_n \}$ is characterized by a smooth $RG$-invariant asymptotic spectral density of the masses squared $\rho(m^{2})$ for large masses, with dimension of the inverse of a mass squared. \par
We define the asymptotic spectral density as follows.
For any test function $f$ the spectral sum can be approximated asymptotically by an integral, keeping the leading term:
\bea
\sum_{n=1}^{\infty} f(m^{2}_n ) \sim \int_{1}^{\infty} f(m^{2}_n ) dn
\eea
in the Euler-MacLaurin formula \cite{Mig1}:
 \begin{equation} \label{EM}
\sum_{k=k_1}^{\infty}G_k(p)=
\int_{k_1}^{\infty}G_k(p)dk - \sum_{j=1}^{\infty}\frac{B_j}{j!}\left[\partial_k^{j-1}G_k(p)\right]_{k=k_1}
\end{equation}
Then by definition the (scalar) asymptotic spectral density satisfies:
\bea
\frac{d n}{d m^{2}}=\rho(m^2)
\eea
i.e. :
\bea
 \int_{1}^{\infty} f(m^{2}_n ) dn =  \int_{ m^{2}_1   }^{\infty}  f(m^2) \rho (m^2) dm^2
 \eea
 Comparing Eq.(\ref{KL}) with the $OPE$ in Eq.(\ref{ope0}), since the scalar condensates behave as:
 \bea
  <\mathcal{O}_{D_1}(0)> \sim Z_{\mathcal{O}_{D_1}}(\frac{\mu}{\Lambda_{QCD}}) \, \Lambda_{QCD}^{D_1} \sim \Lambda_{QCD}^{D_1}
 \eea
 we expand the unknown coefficients  $\mathcal{R}_{n}$ in powers of  $\Lambda_{QCD}$:
 \bea
  \mathcal{R}_{n}=\sum_{D_1=0} \mathcal{R}^{(2D-D_1)}_{n} \Lambda_{QCD}^{D_1}
  \eea
Besides, we write an ansatz for $\mathcal{R}^{(d)}_{n}$:
\bea
\mathcal{R}^{(d)}_{n}= R^{(d)}_n m^{d-4}_n \rho^{-1}(m_n^2)
\eea
with $d=2D-D_1$.
This ansatz in not restrictive and follows only by dimensional analysis to the extent the dimensionless pure numbers $R^{(d)}_n$ are unspecified yet.
However, the specific form of the ansatz is the most convenient for our aims. \par
Substituting the ansatz in Eq.(\ref{KL}) and comparing with the $OPE$ in Eq.(\ref{ope0}), we get the spectral representation for the scalar coefficient functions in the scalar $OPE$:
\bea \label{sp}
C_{d}(x)=  \sum_{n=1}^{\infty}  \frac{1}{(2 \pi)^4} \int \frac{ R^{(d)}_n m^{d-4}_n \rho^{-1}(m_n^2) } {p^2+m^2_n }\,e^{ ip\cdot x}d^4p 
\eea
provided we are able to fix asymptotically $R^{(d)}_n$ for large $n$. In order to do so we express the free propagator in terms of the modified Bessel function $K_1$:
\bea
&& C_{d}(x) = \frac{1}{4 \pi^2 x^2} \sum_{n=1}^{\infty}    R^{(d)}_n m^{d-4}_n \rho^{-1}(m_n^2) \sqrt{x^2 m^2_n} K_1( \sqrt{x^2 m^2_n} ) 
\eea
Now we substitute to the sum the integral, employing the Euler-McLaurin formula. \par
We disregard the terms involving the Bernoulli numbers, because we see from Eq.(\ref{sp}) and Eq.(\ref{EM}) that for each fixed $C_d(x)$ they are suppressed by inverse powers of momentum. \par
However, these power-suppressed terms mix in general in the $OPE$ between different $d$, but they do not carry logarithms, because they arise differentiating the free propagator of lowest mass in the Euler-MacLaurin formula Eq.(\ref{EM}). Their existence requires a short detour in relation to the asymptotic theorem. \par
Indeed, it seems that they introduce power-like suppressed corrections to the perturbative $OPE$, that have not anomalous dimensions \cite{Mig1}. \par
This looks puzzling, and in \cite{Mig1} it is claimed that they must cancel or be absent. Requiring the absence of these power-like corrections is, contrary to the basic philosophy of this paper, an implicit constraint on the actual spectrum \cite{Mig1}. This leads to the approach based on meromorphization and to the representation of two-point connected correlators as ratios of Bessel functions \cite{Mig1}, as opposed to the
Kallen-Lehmann representation employed in this paper.\par
Moreover, the representation by ratios of Bessel functions implies that the spectrum of the masses squared is asymptotically quadratic as shown originally in \cite{Mig}, matching the spectrum occurring in $AdS$-based models of $QCD$-like theories rediscovered (but for different reasons, see \cite{Mig1} and references therein) twenty years later than in \cite{Mig}. \par
Firstly, we observe that the occurrence of power-suppressed terms with no logarithm seems to be an inevitable consequence of the Kallen-Lehmann representation of the two-point correlators as a sum of free propagators, as opposed to the ratio of Bessel functions. Indeed, any sum of a finite number of terms in the series of free propagators gives rise to such power-like suppressed terms with no logarithm, while it is the asymptotics of the series, that depends on an infinite number of terms, that generates the logarithmic behavior. \par
Secondly, for the purposes of the asymptotic solution in this paper, the puzzle is resolved, at least for the first few coefficient functions, that are the most relevant for the asymptotic behavior by dimensional analysis, observing that, precisely 
as a consequence of the absence of logarithms, the first few terms in the expansion in inverse powers of momentum of the terms arising by Bernoulli numbers give rise in fact to contact terms, and therefore can be ignored. \par
Indeed, for an operator of large naive dimension, they are multiplied by large positive powers of momentum, as it becomes clear working out the asymptotic theorem in the
momentum representation in Eq.(\ref{mom}).\par
For example, in the $OPE$ of $TrF^2$ we get for the first-two coefficient functions in the momentum representation \cite{MBN,MBM,MBFloer}, up to contact terms:
\bea \label{mope1}
C_8(p) && \sim p^4  \big(g^2(p) + O(\frac{\Lambda_{QCD}^2}{p^2+\Lambda^2_{QCD}})\big) 
\eea
\bea \label{mope2}
C_4(p) && \sim g^2(p) + O(\frac{\Lambda_{QCD}^2}{p^2+\Lambda^2_{QCD}})
\eea
with $g(p)$ the running coupling, and for convenience we have chosen a scheme in which the mass gap in  the scalar glueball sector coincides with $\Lambda_{QCD}$. In both cases the first term involves logarithms, while the second term involves Bernoulli numbers, is power-suppressed, and has no logarithm. \par 
Expanding the lowest-mass propagator for large momentum, we get:
\bea \label{mope11}
C_8(p) && \sim p^4  \big(g^2(p) + O(\frac{\Lambda_{QCD}^2}{p^2}) +O(\frac{\Lambda_{QCD}^4}{p^4})+O(\frac{\Lambda_{QCD}^6}{p^6}) + \cdots\big) \nonumber \\
&& \sim  p^4  g^2(p) + O(p^2 \Lambda_{QCD}^2) +O(\Lambda_{QCD}^4)+O(\frac{\Lambda_{QCD}^6}{p^2}) + \cdots
\eea
$C_8(p)$ and $C_4(p)$ mix together in the two-point correlator of $TrF^2$ because of Eq.(\ref{ope0}):
\bea
G^{(2)}(p) \sim C_8(p) + C_4(p) \Lambda_{QCD}^4+ \cdots
\eea
Now the first-two power-suppressed terms
in Eq.(\ref{mope11}) are larger than the first term in Eq.(\ref{mope2}), that is logarithmically suppressed, thus spoiling seemingly the asymptotic estimate for $C_4(p)$ implied by the $OPE$,
because of the mixing in Eq.(\ref{ope0}) with these larger terms.\par
But in fact these two larger terms, precisely because they do not carry logarithms, are polynomial in momentum, i.e. contact terms, i.e. irrelevant. Therefore, the universal logarithmic behavior of $C_4(p)$ implied by the $OPE$ is not spoiled.\par
In the same fashion, the universal logarithmic behavior of the coefficient functions $C_{D}(p)$ implied by the $OPE$ for any sufficiently large $D$ is not spoiled by the power-suppressed corrections to $C_{2D}(p),C_{2D-2}(p), \cdots $ associated to Bernoulli numbers, independently on the final fate of the power-suppressed corrections, i.e. whether they cancel or not in the exact solution according to the point of view in \cite{Mig1}. \par
This happens for all the coefficients functions in the $OPE$ that have non-negative naive dimension in the momentum representation. This suffices for the aims of the asymptotic solution in the scalar sector. The spin-$1$
sector for fermion bilinears of naive dimension $D=3$ is a special case considered at the end of this section.   \par
Coming back to our computation in the generic case, we get asymptotically for the leading contribution for each $d$:
\bea
C_d(x)
&& \sim \frac{1}{4 \pi^2 x^2} \int_{1}^{\infty}    R^{(d)}_n m^{d-4}_n \rho^{-1}(m_n^2) \sqrt{x^2 m^2_n} K_1( \sqrt{x^2 m^2_n} ) dn \nonumber \\
&& = \frac{1}{4 \pi^2 x^2} \int_{m^2_1}^{\infty}   R^{(d)}(m)  m^{d-4}  \sqrt{x^2 m^2} K_1( \sqrt{x^2 m^2} ) dm^2
\eea
because in the generic case the terms arising by Bernoulli numbers are power suppressed. 
We notice that the spectral distribution $\rho(m^2)$ cancels thanks to the convenient choice in our ansatz. This feature allows us to perform explicit computations, and it shows why the asymptotic solution does not contain spectral information. \par
We introduce now the dimensionless variable $z^2=x^2 m^2$:
\bea \label{R0}
C_d(x) && \sim \frac{1}{4 \pi^2 x^2} \int_{m^2_1}^{\infty}    R^{(d)}(m) m^{d-4}  \sqrt{x^2 m^2} K_1( \sqrt{x^2 m^2} ) dm^2 \nonumber \\
&& = \frac{1}{4 \pi^2 x^2} \int_{m^2_1 x^2}^{\infty}    R^{(d)}(\frac{z}{x}) (\frac{z}{x})^{d-4}  z K_1(z) \frac{dz^2}{x^2} \nonumber \\
&& = \frac{1}{4 \pi^2 x^{d}} \int_{m^2_1 x^2}^{\infty}    R^{(d)}(\frac{z}{x}) z^{d-3}   K_1(z) dz^2
\eea
with $x=\sqrt{x^2}$. \par
Because of the universal asymptotic behavior of the running coupling $g(x)$:
\bea
g^2(x)\sim \frac{1}{\beta_0\log(\frac{1}{x^2 \Lambda^2_{QCD}})}\biggl(1-\frac{\beta_1}{\beta_0^2}\frac{\log\log(\frac{1}{x^2 \Lambda^2_{QCD}})}{\log(\frac{1}{x^2 \Lambda^2_{QCD}})}\biggr)
\eea
it follows from Eq.(\ref{R0}) and Eq.(\ref{acf}) that it must hold:
\bea \label{eq}
\int_{m_1^2 x^2}^{\infty}    R^{(d)}(\frac{z}{x}) z^{d-3}   K_1(z) dz^2 \sim \Biggl(\frac{1}{\beta_0\log(\frac{1}{x^2 \Lambda^2_{QCD}})}\biggl(1-\frac{\beta_1}{\beta_0^2}\frac{\log\log(\frac{1}{x^2 \Lambda^2_{QCD}})}{\log(\frac{1}{x^2 \Lambda^2_{QCD}})}\biggr)\Biggr)^{\frac{\gamma_{0d}}{\beta_0}+l}
\eea
with $\gamma_{0d}= \gamma_0(\mathcal{O}_D)  - \frac{1}{2}\gamma_0(\mathcal{O}_{D_1})$.
For the applications to the scalar case in this paper it suffices to restrict to $d >2$. The borderline case $d=2$ is special and it is relevant for the coefficient function of a spin-$1$ current constructed by a fermion bilinear of dimension $D=3$ in the $OPE$ of two currents. It will be considered at the end of this section. \par
The asymptotic solution for $d >2$ is:
\bea \label{sol1}
R^{(d)}(\frac{z_0}{x}) \sim \Biggl(\frac{1}{\beta_0\log(\frac{z_0^2}{x^2 \Lambda^2_{QCD}})}\biggl(1-\frac{\beta_1}{\beta_0^2}\frac{\log\log(\frac{z_0^2}{x^2 \Lambda^2_{QCD}})}{\log(\frac{z_0^2}{x^2 \Lambda^2_{QCD}})}\biggr)\Biggr)^{\frac{\gamma_{0d}}{\beta_0}+l}
\eea
or equivalently, within the universal, i.e. scheme independent, asymptotic accuracy:
\bea  \label{sol2}
R^{(d)}(\frac{z_0}{x}) \sim  g(x)^{2l} Z^2_{\mathcal{O}_D}(x \mu ,g(\mu)) Z^{-1}_{\mathcal{O}_{D_1}}(x \mu ,g(\mu)) 
\eea
Indeed, the universal part of $R^{(d)}(\frac{z_0}{x})$ is actually $z_0$ independent, because a rescaling of $\frac{1}{x}$ by any fixed $z_0$ is equivalent to a change of scheme, i.e. a rescaling by $\frac{1}{z_0}$ of $\Lambda_{QCD}$ in Eq.(\ref{sol1}). It may seem that the roles of $\frac{1}{x}$ and of $z$ are actually symmetric in this argument, but they are not, because the variable $z$ is integrated over, and the corresponding integral is convergent due to the condition $d>2$, while we look for the asymptotic behavior in the variable $x$, that is not integrated over.
Therefore, by the dominated convergence theorem, we can put $R^{(d)}(\frac{z}{x})$, for any fixed $z=z_0$, say $z_0=1$, outside the integral over $z$ asymptotically in the limit $x \rightarrow 0$:
\bea
&&\int_{m_1^2 x^2}^{\infty}    R^{(d)}(\frac{z}{x}) z^{d-3}   K_1(z) dz^2 \sim R^{(d)}(\frac{1}{x}) \int_{0}^{\infty}    z^{d-3}   K_1(z) dz^2 \nonumber \\
&&\sim \Biggl(\frac{1}{\beta_0\log(\frac{1}{x^2 \Lambda^2_{QCD}})}\biggl(1-\frac{\beta_1}{\beta_0^2}\frac{\log\log(\frac{1}{x^2 \Lambda^2_{QCD}})}{\log(\frac{1}{x^2 \Lambda^2_{QCD}})}\biggr)\Biggr)^{\frac{\gamma_{0d}}{\beta_0}+l}
\eea
since the integrals:
\bea
\int_{0}^{\infty}    z^{d-3}   K_1(z) dz^2 
\eea
and
\bea
\int_{0}^{\infty}     R^{(d)}(\frac{z}{x})z^{d-3}   K_1(z) dz^2
\eea
and all the integrals in $dz^2$ of the terms that depend on $z$ obtained in the expansion of the ansatz in Eq.(\ref{sol1}) for $R(\frac{z}{x})$ by means of:
\bea
\frac{1}{\log(\frac{z^2}{x^2 \Lambda^2_{QCD}})}=\frac{1}{\log(\frac{1}{x^2 \Lambda^2_{QCD}})}(1+\frac{\log(z^2)}{\log(\frac{1}{x^2 \Lambda^2_{QCD}})})^{-1}
\eea
are convergent for $d>2$, because $K_1$ has a simple pole in $z=0$ and decays exponentially for large $z$. Incidentally, this shows also that the series in Eq.(\ref{sp}) is convergent for $x\neq 0$. \par
Eq.(\ref{sol1}) and Eq.(\ref{sol2}) have two main applications in section \ref{s2}. \par
For $d=2D$ they reduce to $ R^{(2D)}_n \sim Z^2_{n}$, since $l=0$ because $C_2 \neq 0$, that gives us the asymptotic spectral representation of the two-point correlator in Eq.(\ref{002}):
\bea
G^{(2)}(x) \sim C_{2D}(x) \sim \sum_{n=1}^{\infty}  \frac{1}{(2 \pi)^4} \int  \frac{ m^{2D-4}_n Z_n^2 \rho^{-1}(m_n^2) }{p^2+m^{ 2}_{n}} \,e^{ ip\cdot x}d^4p   
\eea
In this case the residues at the poles are necessarily positive, because this contribution is the leading contribution to the two-point correlator, in such a way that the spectral sum converges monotonically when it is convergent. Thus the residues cannot oscillate in sign and the Euler-MacLaurin formula is justified. In fact, the Euler-MacLaurin formula needs some sort of regularity assumptions to be mathematically completely justified. \par
For $d=D$ Eq.(\ref{sol1}) and Eq.(\ref{sol2}) reduce to $ R^{(D)}_n \sim Z_{n}$, since $l=0$ because $C_3 \neq 0$ by assumption, that gives us the asymptotic spectral representation of the coefficient function of the operator $\mathcal{O}$ itself in Eq.(\ref{001}):
\bea \label{osc}
C(x) \sim C_{D}(x) \sim \sum_{n=1}^{\infty}  \frac{1}{(2 \pi)^4} \int  \frac{ m^{D-4}_n Z_n \rho^{-1}(m_n^2) }{p^2+m^{ 2}_{n}} \,e^{ ip\cdot x}d^4p   
\eea
In this case the residues at the poles need not to be positive, and from a purely mathematical point of view there is the possibility that they oscillate in sign wildly, i.e. at the highest frequency possible: They may change sign for any new addendum in the spectral sum. \par
If this is the case the solution obtained by the Euler-Maclaurin formula may not be the asymptotically unique solution. More precisely, the Euler-MacLaurin formula leads to an integral equation of Fredholm type  in
the momentum representation for which the solution is unique if and only if it exists by the Fredholm alternative (see remark below Eq.(\ref{15})). But the Euler-MacLaurin formula may not apply strictly because the oscillating part can sum to a subleading contribution due to cancellations in the alternating sum, yet the individual residues may not be subleading. \par
Neverthless, in the generic case \footnote{I.e. for operators with large dimension with respect to the twist.} even in presence of additional wildly-oscillating contributions not taken into account by the Euler-MacLaurin formula, the solution to the asymptotic behavior obtained by the Euler-MacLaurin formula captures necessarily the leading asymptotic behavior of the correlators and therefore of the asymptotic inclusive $S$-matrix restricted to sectors of fixed spin, that involve the sum over all final states of given spin, realizing in any case a precise version of the gluon-quark/glueball-meson (asymptotic) duality, that overall is the real physical input of the asymptotically-free bootstrap. \par
The asymptotic theorem in the coordinate representation extends to the two-point correlators of integer spin $s$:
 \bea
 G_s^{(2)}(x)= \langle \mathcal{O}^{(s)}(x) \mathcal{O}^{(s)}(0) \rangle_{conn}\, 
 \eea
after a short detour in the momentum representation. \par
At first this seems just a completely trivial change, since it is obtained by Fourier transforming. Yet a new complication arises
in the momentum representation. 
From now on we restrict to $d$ even \footnote{This is the case, for example, for the coefficient function of the identity operator in the $OPE$ of any operator that creates from the vacuum glueballs or mesons, and for all the scalar coefficient functions in the scalar $OPE$ in the pure-glue scalar sector
of large-$N$ $QCD$, because all such scalar operators have naive dimension even.},
because the sum over one-particle states:
\bea \label{spm}
C_{d}(p) \sim  \sum_{n=1}^{\infty}  \frac{ R^{(d)}_n m^{d-4}_n \rho^{-1}(m_n^2) } {p^2+m^2_n }
\eea
 is always divergent for $d>4$ due to the grow of the residues, 
but for $d > 4$ even the divergence is actually a polynomial in momentum of finite degree but with infinite coefficients, i.e. an infinite contact term. \par
In order to prove this statement we use the identity:
\bea \label{b1}
m_n^{d-4} = (-p^2+p^2+m_n^2)^{\frac{d}{2}-2}
\eea
from which we get:
\bea \label{b2}
C_d(p) \sim (- p^2)^{\frac{d}{2}-2} \sum_{n=1}^{\infty} \frac{R_n  \rho^{-1}(m_n^2) }{p^2+m^2_n } + \cdots
\eea
where the dots represent contact terms, i.e. distributions whose Fourier transform is supported at $x=0$, that are physically irrelevant and that therefore can be safely discarded. \par
The contact terms arise because, for $\frac{d}{2}$ even and $\frac{d}{2}-2$ non-negative integer, in Eq.(\ref{b2}) in addition to the term $(- p^2)^{\frac{d}{2}-2}$ at least one term involving the factor of $p^2+m^{2}_n$, that cancels the denominator, always occurs. \par
It is also clear from Eq.(\ref{b2}) and Eq.(\ref{b1}) that the sum of the contact terms is divergent, but nevertheless the entire sum, and not just the individual terms, is a polynomial of finite degree in momentum. \par
Besides, for higher spins by the Kallen-Lehmann representation:
\bea \label{mom}
G_s^{(2)}(p)&&=  \sum_{n=1}^{\infty}  P^{(s)} \big(\frac{p_{A}}{m^{(s)}_n}\big) \frac{m^{(s)2D-4}_n R_n^{(s)}  \rho_s^{-1}(m^{(s)2}_n)}{p^2+m^{(s)2}_n  }    \nonumber \\
&&=P^{(s)} \big(\frac{p_{A}}{p} \big)  \, p^{2D-4}   \sum_{n=1}^{\infty} \frac{R_n^{(s)}   \rho_s^{-1}(m^{(s)2}_n)  }{p^2+m^{(s)2}_n  }
     + \cdots
\eea
for some $R_n^{(s)} $ to be fixed asymptotically momentarily,
where the dots represent contact terms and $P^{(s)} \big(\frac{p_{A}}{p} \big)$ is the projector obtained substituting $-p^2$  to $m_n^2$ in  $P^{(s)} \big(\frac{p_{A}}{m_n} \big)$. \par
The proof is as follows.
$m^{(s)2D-4}_n P^{(s)} \big(\frac{p_{A}}{m^{(s)}_n}\big)$ is a polynomial in powers of $m^2_n$. To each monomial $m^{2d}_n$ occurring in this polynomial we can substitute either $p^{2d}$ or $- p^{2d}$, for $d$ even or $d$ odd respectively,
up to contact terms, because of Eq.(\ref{b2}). This is the same as substituting  $-p^2$  to $m_n^2$ in  $P^{(s)} \big(\frac{p_{A}}{m_n} \big)$ since for $d$ even we always get a positive sign.\par
Now going back to the coordinate representation after subtracting contact terms:
\bea \label{coo}
G_s^{(2)}(x)=P^{(s,2D-4)} \big(\frac{\partial}{\partial x_A} \big)  \,    \sum_{n=1}^{\infty}  \frac{1}{(2 \pi)^4} \int \frac{R_n^{(s)}   \rho_s^{-1}(m^{(s)2}_n)  }{p^2+m^{(s)2}_n  } e^{ip\cdot x}d^4p   \nonumber \\
\eea
with $P^{(s,2D-4)}$ a homogeneous polynomial of degree $2D-4$ in the derivatives. If we set $R_n^{(s)}=Z_n^{(s)2}$ then:
\bea \label{coo1}
G_s^{(2)}(x)=P^{(s,2D-4)} \big(\frac{\partial}{\partial x_A} \big)  \,   \sum_{n=1}^{\infty}  \frac{1}{(2 \pi)^4} \int \frac{Z_n^{(s)2}   \rho_s^{-1}(m^{(s)2}_n)  }{p^2+m^{(s)2}_n  } e^{ip\cdot x}d^4p  \nonumber \\
\eea
satisfies asymptotically the correct Callan-Symanzik equation:
\begin{equation}\label{a}
\left(x_{M}\frac{\partial}{\partial x_{M}}+\beta(g)\frac{\partial}{\partial g}+2(D+\gamma_{\mathcal{O}^{(s)}}(g))\right) G_s^{(2)}(x)  =0 
\end{equation}
since by the scalar version of the asymptotic theorem the sum:
\bea
S(x)=\sum_{n=1}^{\infty}  \frac{1}{(2 \pi)^4} \int \frac{Z_n^{(s)2}   \rho_s^{-1}(m^{(s)2}_n)  }{p^2+m^{(s)2}_n  } e^{ip\cdot x}d^4p
\eea
provided it is convergent (the conditions for convergence will be stated momentarily), satisfies asymptotically:
\begin{equation}\label{aa}
\left(x_{M}\frac{\partial}{\partial x_{M}}+\beta(g)\frac{\partial}{\partial g}+2(2+\gamma_{\mathcal{O}^{(s)}}(g))\right) S(x)  =0
\end{equation}
that follows from $R_n^{(s)}=Z_n^{(s)2}$ and the previous analysis in the scalar case. Hence the asymptotic theorem extends to any integer spin.\par
Once the (divergent) contact terms have been subtracted, it is possible to obtain the asymptotic behavior directly in the momentum representation. \par
Applying again the Euler-MacLaurin formula:
\bea \label{15}
 C_d(p) && \sim p^{d-4}
\int_{1}^{\infty} \frac{R^{(d)}_n  \rho^{-1}(m_n^2) }{p^2+m^2_n } dn \nonumber \\
&&= p^{d-4}\int_{m^2_1}^{\infty} \frac{R^{(d)}(m)  \rho^{-1}(m^2) }{p^2+m^2 } \rho(m^2) dm^2 \nonumber \\
&&= p^{d-4}\int_{m^2_1}^{\infty} \frac{R^{(d)}(m) }{p^2+m^2 } dm^2\nonumber \\
&& \sim p^{d-4} I^{(d)} ( \frac{p^2} {\Lambda_{QCD}^2} )
 \eea
 where we have set for convenience $m_1=\Lambda_{QCD}$.
This is an integral equation of Fredholm type (after introducing eventually an $UV$ cutoff below), whose solution by the Fredholm alternative
exists if and only if it is unique \cite{MBN}. \par
The unique asymptotic solution for $R^{(d)}(m)$ is given by:
\bea \label{sol11}
R^{(d)}(m) \sim \Biggl(\frac{1}{\beta_0\log(\frac{m^2}{ c \Lambda^2_{QCD}})}\biggl(1-\frac{\beta_1}{\beta_0^2}\frac{\log\log(\frac{m^2}{c \Lambda^2_{QCD}})}{\log(\frac{m^2}{c \Lambda^2_{QCD}})}\biggr)\Biggr)^{\frac{\gamma_{0d}}{\beta_0}+l}
\eea
because of Eq.(\ref{sol1}), with the constant $c$ scheme dependent, as we check momentarily independently on Eq.(\ref{sol1}). Introducing the dimensionless variables $\nu=\frac{p^2}{\Lambda_{QCD}^2}$ and $k=\frac{m^2}{\Lambda_{QCD}^2}$, 
we distinguish two cases: Either $\gamma'=\frac{\gamma_{0d}}{\beta_0}+l >1$ or otherwise. \par
We get for $I^{(d)}(\nu)$:
\begin{align} \label{14}
I^{(d)}(\nu)& \sim \int_1^{\infty}\beta_0^{-\gamma'}\left(\frac{1}{\log(\frac{k}{c})}\left(1-\frac{\beta_1}{\beta_0^2}\frac{\log\log(\frac{k}{c})}{\log(\frac{k}{c})}\right)\right)^{\gamma'}\frac{dk}{k+\nu}\nonumber\\
&=\beta_0^{-\gamma'}\int_{1+\nu}^{\infty}\left(\frac{1}{\log(\frac{k-\nu}{c})}\left(1-\frac{\beta_1}{\beta_0^2}\frac{\log\log(\frac{k-\nu}{c})}{\log(\frac{k-\nu}{c})}\right)\right)^{\gamma'}\frac{dk}{k}\nonumber\\
&\sim \beta_0^{-\gamma'}\int_{1+\nu}^{\infty}\left[\log(\frac{k-\nu}{c})\right]^{-\gamma'}
\left(1-\gamma'\frac{\beta_1}{\beta_0^2}\frac{\log\log(\frac{k-\nu}{c})}{\log(\frac{k-\nu}{c})}\right)\frac{dk}{k}\nonumber\\
&\sim  \beta_0^{-\gamma'}\int_{1+\nu}^{\infty}\left[\log(\frac{k-\nu}{c})\right]^{-\gamma'}\frac{dk}{k} 
-\gamma'\frac{\beta_1}{\beta_0^2}\beta_0^{-\gamma'}\int_{1+\nu}^{+\infty}\left[\log(\frac{k-\nu}{c})\right]^{-\gamma'-1}\log\log(\frac{k-\nu}{c})
\frac{dk}{k}
\end{align}
where following \cite{MBN,MBM}  we have changed variables in the $RHS$ of Eq.(\ref{14}) from $k$ to $k+\nu$. \par
For $\gamma'>1$
the integral in Eq.(\ref{14}) is convergent, in such a way that the integration domain can be extended to $\infty$. Otherwise, the integral is divergent, but the divergence is a contact term. Therefore, after subtracting the contact term, the integration domain can be extended to $\infty$ anyway. \par
For the first integral in the last line we get:
\begin{align}\label{eqn: int1}
&\int_{1+\nu}^\infty\frac{1}{k}[\beta_0\log(\frac{k}{c})]^{-\gamma'}
\biggl[1+\frac{\log(1-\frac{\nu}{k})}{\log(\frac{k}{c})}\biggr]^{-\gamma'}dk\nonumber\\
&\sim \int_{1+\nu}^\infty\frac{1}{k}[\beta_0\log(\frac{k}{c})]^{-\gamma'} 
\biggl[1+\gamma'\frac{\nu}{k \log(\frac{k}{c})}\biggr]dk\nonumber\\
&=\int_{1+\nu}^\infty\frac{1}{k}[\beta_0\log(\frac{k}{c})]^{-\gamma'}dk+
\gamma' \nu\int_{1+\nu}^\infty\frac{1}{k^2}\beta_0^{-\gamma'}[\log(\frac{k}{c})]^{-\gamma'-1}dk 
\end{align}
It follows the leading asymptotic behavior, provided $\gamma' \neq 1$:
\begin{equation}\label{eqn:sol_int_leading}
\int_{1+\nu}^\infty\frac{1}{k}[\beta_0\log(\frac{k}{c})]^{-\gamma'}dk=
\frac{1}{\gamma'-1} \beta_0^{-\gamma'}\left[\log\left(\frac{1+\nu}{c}\right)\right]^{-\gamma'+1}
\end{equation}
For $\gamma' = 0$ there is nothing to add. It corresponds to the asymptotically-conformal case in the $UV$. We leave the case $\gamma' = 1$ as an easy exercise. If $\gamma' \neq 0,1$ we
add the second contribution. We evaluate it at the leading order by changing variables and integrating by parts:
\begin{align}\label{eqn:int_next-to-leading}
&\gamma'\frac{\beta_1}{\beta_0^2}\beta_0^{-\gamma'}\int_{1+\nu}^{+\infty}\left[\log(\frac{k-\nu}{c})\right]^{-\gamma'-1}\log\log(\frac{k-\nu}{c})
\frac{dk}{k}\nonumber\\
&\sim  \gamma'\frac{\beta_1}{\beta_0^2}\beta_0^{-\gamma'}\int_{1+\nu}^{+\infty}\left[\log(\frac{k}{c})\right]^{-\gamma'-1}\log\log(\frac{k}{c})
\frac{dk}{k}\nonumber\\
&=\gamma'\frac{\beta_1}{\beta_0^2}\beta_0^{-\gamma'}\int_{\log\frac{1+\nu}{c}}^{+\infty}t^{-\gamma'-1}\log(t)dt\nonumber\\
&=\gamma'\frac{\beta_1}{\beta_0^2}\beta_0^{-\gamma'}\left[\frac{1}{\gamma'}\left(\log(\frac{1+\nu}{c})\right)^{-\gamma'}\log\log(\frac{1+\nu}{c})+
\frac{1}{\gamma'^2}\left(\log(\frac{1+\nu}{c})\right)^{-\gamma'}\right]
\end{align}
The second term in brackets in the last line is subleading with respect to the first one. 
Collecting Eq.(\ref{eqn:int_next-to-leading}) and Eq.(\ref{eqn:sol_int_leading}) we get for $I^{(d)}(\nu)$:
\begin{align}\label{eqn:esp_ntl}
I^{(d)}(\nu)
& \sim \frac{1}{\gamma'-1}\beta_0^{-\gamma'}\left(\log\frac{1+\nu}{c}\right)^{-\gamma'+1}-\frac{\beta_1}{\beta_0^2}\beta_0^{-\gamma'}\left(\log(\frac{1+\nu}{c})\right)^{-\gamma'}\log\log(\frac{1+\nu}{c})\nonumber\\
&=\frac{\beta_0^{-\gamma'}}{\gamma'-1}\biggl(\log\frac{1+\nu}{c}\biggr)^{-\gamma'+1}\left[1-\frac{\beta_1(\gamma'-1)}{\beta_0^2}\left(\log(\frac{1+\nu}{c})\right)^{-1}\log\log(\frac{1+\nu}{c})\right]\nonumber\\
&\sim \frac{1}{\beta_0(\gamma'-1)}\left(\beta_0\log\frac{1+\nu}{c}\right)^{-\gamma'+1}\left[1-\frac{\beta_1}{\beta_0^2}\left(\log(\frac{1+\nu}{c})\right)^{-1}\log\log(\frac{1+\nu}{c})\right]^{\gamma'-1}\nonumber\\
&\sim \Biggl(\frac{1}{\beta_0\log \nu}\biggl(1-\frac{\beta_1}{\beta_0^2}\frac{\log\log\nu}{\log\nu}\biggr)\Biggr)^{\gamma'-1}
\end{align}
Hence we get the asymptotic theorem for two-point correlators in the momentum representation:
\bea \label{CS}
G_s^{(2)}(p)
 \sim P^{(s)}\big(\frac{p_{A}}{p}\big) \, p^{2D-4}    \Biggl[\frac{1}{\beta_0\log (\frac{p^2}{\Lambda^2_{QCD} } )}\biggl(1-\frac{\beta_1}{\beta_0^2}\frac{\log\log (\frac{p^2}{\Lambda^2_{QCD} } ) }{\log (\frac{p^2}{\Lambda^2_{QCD} } )}      \biggr)\Biggr]^{\gamma'-1} \nonumber \\
\eea
up to contact terms, with $l=0$ and $\gamma' \neq 0,1$. \par
Because of the applications, we write explicitly the spin-$1$ and the spin-$2$ two-point correlators and the coefficient functions of the spin-$1$ currents themselves, that still require a detailed study. \par
We employ Veltman conventions for Euclidean and Minkowskian propagators  (see Appendix F in \cite{Velt2}). \par
For spin $1$ in vector notation:
\bea
&& \int \langle \mathcal{O}^{(1)}_{A}(x) \mathcal{O}^{(1)}_{B}(0) \rangle_{conn}\,e^{-ip\cdot x}d^4x  \nonumber \\
&& \sim \sum_{n=1}^{\infty} (\delta_{AB } +  \frac{p_{A} p_{B}}{m^{(1) 2}_n}) \frac{m^{(1)2D-4}_n Z_n^{(1)2}  \rho_1^{-1}(m^{(1)2}_n)}{p^2+m^{(1)2}_n  } \nonumber \\
&& \sim  \, p^{2D-4}     (\delta_{AB} -  \frac{p_{A} p_{B}}{p^2}) \sum_{n=1}^{\infty}   \frac{Z_n^{(1)2}   \rho_1^{-1}(m^{(1)2}_n)  }{p^2+m^{(1)2}_n  }
     + \cdots
\eea
For spin $2$ in vector notation:
\bea
&& \int \langle \mathcal{O}^{(2)}_{AB}(x) \mathcal{O}^{(2)}_{CD}(0) \rangle_{conn}\,e^{-ip\cdot x}d^4x  \sim \nonumber \\
&&  \sum_{n=1}^{\infty} \bigg[\frac{1}{2} \eta_{A C}(m^{(2)}_n) \eta_{B D}(m^{(2)}_n) + \frac{1}{2} \eta_{B C}(m^{(2)}_n)  \eta_{A D}(m^{(2)}_n) - \frac{1}{3} \eta_{A B}(m^{(2)}_n)  \eta_{C D}(m^{(2)}_n) \bigg]  \frac{m^{(2)2D-4}_n Z_n^{(2)2}  \rho_2^{-1}(m^{(2)2}_n)}{p^2+m^{(2)2}_n  } \nonumber \\
&&\sim  \, p^{2D-4}  \bigg[\frac{1}{2}  \eta_{A B}(p) \eta_{B C}(p) + \frac{1}{2} \eta_{B C}(p)  \eta_{A D}(p) - \frac{1}{3} \eta_{A B}(p)  \eta_{CD}(p) \bigg]     \sum_{n=1}^{\infty}   \frac{Z_n^{(2)2}   \rho_2^{-1}(m^{(2)2}_n)  }{p^2+m^{(2)2}_n  }+\cdots
\eea
where:
\bea
\eta_{A B}(m)= \delta_{A B} +  \frac{p_{A} p_{B}}{m^2}
\eea
and:
\bea
\eta_{AB}(p)= \delta_{AB} -  \frac{p_{A} p_{B}}{p^2}
\eea
Some observations are in order. \par
For the vector and axial currents in $QCD$, the gluino chiral currents in $\mathcal{N}=1$ $SUSY$ $YM$, and the stress energy tensor, $\lim_{n \rightarrow \infty} Z_n \rightarrow 1$ because of current conservation at large-$N$. \par
Each massive propagator is conserved only on the respective mass shell. However, after subtracting the sum of contact terms denoted by the dots, the resulting massless projector implies off-shell conservation, as if the large-$N$ $QCD$ propagators
were saturated by massless particles only. This is necessary to match $QCD$ perturbation theory (with massless quarks). For a direct check see \cite{chet,chet2}. \par
For higher spins, it is possible to write down local actions reproducing the two-point massive correlators, introducing eventually auxiliary fields (see \cite{Francia1,Francia2} and references therein). \par
In the spin-$2$ case the massless projector contains a factor of $\frac{1}{3}$ in the last term, that descends from the massive case, and not of $\frac{1}{2}$, that would occur for a truly physical massless spin-$2$ particle according to Van Dam-Veltman-Zakharov  discontinuity \cite{Velt,Zak}. This factor of $\frac{1}{3}$ occurs also in perturbative computations of the correlator of the stress-energy tensor in $QCD$ \cite{chet}. Indeed,
a spin-$2$ glueball in $QCD$ is not a graviton. \par
We now study the coefficient function in the $OPE$ of the current itself: 
\bea \label{CF}
j^{a\alpha_1 \dot \beta_1}_{R }(x) j^{b\alpha_2\dot \beta_2}_{R  }(0) = C^{ab\alpha_1 \dot \beta_1 \alpha_2 \dot \beta_2 } _{R c' \alpha' \dot \beta'}(x) j^{c' \alpha' \dot \beta'}_{R }(0) + \cdots
\eea
For example, for vector currents in the conformal limit \cite{Todorov}:
\bea
&& <j^{a\alpha_1 \dot \beta_1}_{V}(x_1) j^{b\alpha_2 \dot \beta_2}_{V }(x_2) j^{c\alpha_3 \dot \beta_3}_{V }(x_3)>  \nonumber \\
&&\sim f^{abc}  (\frac{x^{\alpha_1 \dot \beta_2}_{12}} {x_{12}^4}   \frac{x^{\alpha_2 \dot \beta_3}_{23}}{x_{23}^4}\frac{x^{\alpha_3 \dot \beta_1}_{31}}{x_{31}^4}
-  \frac{x^{\alpha_1 \dot \beta_3}_{13}} {x_{13}^4}   \frac{x^{\alpha_3 \dot \beta_2}_{32}}{x_{32}^4}\frac{x^{\alpha_2 \dot \beta_1}_{21}}{x_{21}^4}) \nonumber \\
\eea
that becomes for $x_1 \rightarrow x_2$:
\bea
&& <j^{a\alpha_1 \dot \beta_1}_{V}(x_1) j^{b\alpha_2 \dot \beta_2}_{V }(x_2) j^{c\alpha_3 \dot \beta_3}_{V }(x_3)>  \nonumber \\
&&\sim f^{abc'}  (\frac{x^{\alpha_1 \dot \beta_2}_{12}} {x_{12}^4} <j^{c'\alpha_2 \dot \beta_1}_{V }(x_1) j^{c\alpha_3 \dot \beta_3}_{V }(x_3)> - \frac{x^{\alpha_2 \dot \beta_1}_{21}}{x_{21}^4}
  <j^{c'\alpha_1 \dot \beta_2}_{V} (x_1) j^{c\alpha_3 \dot \beta_3}_{V }(x_3)>) \nonumber \\
\eea
Hence we get in the conformal limit:
\bea
C^{ab\alpha_1 \dot \beta_1 \alpha_2 \dot \beta_2} _{V c' \alpha' \dot \beta'}(x)  \sim f^{abc'} (\frac{x^{\alpha_1 \dot \beta_2}_{12}} {x_{12}^4}       \delta^{\alpha_2}_{ \alpha'}\delta^{\dot \beta_1}_{ \dot \beta'}
- \frac{x^{\alpha_2 \dot \beta_1}_{21}}{x_{21}^4}  \delta^{\alpha_1}_{ \alpha'} \delta^{\dot \beta_2}_{ \dot \beta'})
\eea 
From the definition of the coefficient function in Eq.(\ref{CF}) it follows a Kallen-Lehmann representation for the latter, analog to the two-point function \footnote{In fact, there are additional complications that we ignore, because we are interested only in the large-momentum asymptotic behavior.}, obtained taking matrix elements between the vacuum and a spin-$1$ state in
Eq.(\ref{CF}) and inserting a complete basis of single-particle states
created from the vacuum by the current.
Asymptotically, to match the conformal limit, it reads:
\bea
C^{ab\alpha_1 \dot \beta_1 \alpha_2 \dot \beta_2 }_{V c' \alpha' \dot \beta'}(p)  \sim f^{abc'} (p^{\alpha_1 \dot \beta_2} \delta^{\alpha_2}_{ \alpha'}\delta^{\dot \beta_1}_{ \dot \beta'}
+ p^{\alpha_2 \dot \beta_1} \delta^{\alpha_1}_{ \alpha'} \delta^{\dot \beta_2}_{ \dot \beta'}) \sum_{n=1}^{\infty}  \frac{ m^{-2}_{Vn} R^{(2)}_{Vn}  \rho^{-1}_{1}(m^2_{Vn})}{p^2+m^{2}_{Vn }  } \nonumber \\
\eea 
with:
\bea
 \sum_{n=1}^{\infty}  \frac{1}{(2 \pi)^4} \int  \frac{ m^{-2}_{Vn} R^{(2)}_{Vn}  \rho^{-1}_{1}(m^2_{Vn})}{p^2+m^{2}_{Vn }  }  e^{ip \cdot x} d^4p \sim C_{2V}(x) \sim \frac{1}{4 \pi^2 x^2}
\eea
 for $d=2$, $\gamma_{0d}=l=0$. Since $d=2$, this is the special borderline case not considered previously. This case is borderline also in the sense that the first term arising by Bernoulli numbers is on the same order of the entire spectral sum, because the coefficient function is of negative dimension in the momentum representation. \par
 We get the asymptotic equation:
\bea
&& C_{2V}(x) = \frac{1}{4 \pi^2 x^2} \sum_{n=1}^{\infty}    R^{(2)}_{Vn} m^{-2}_{Vn} \rho^{-1}(m_n^2) \sqrt{x^2 m^2_n} K_1( \sqrt{x^2 m^2_n} ) \sim \frac{1}{4 \pi^2 x^2} \nonumber \\
\eea
Taking the point-like limit $x \rightarrow 0$, since the pole of the modified Bessel function is cancelled by the factor in front of it, it follows that the series:
\bea \label{V}
\sum_{n=1}^{\infty}    R^{(2)}_{Vn} m^{-2}_{Vn} \rho_{1}^{-1}(m_{Vn}^2) \sim 1
\eea
 must be convergent (and non-vanishing). It turns out that it is not convenient to assume that the series in Eq.(\ref{V}) is absolutely convergent, that would allow us to substitute to the sum the integral
 asymptotically.\par
Analogously, for the $R$ currents:
\bea
 C_{2R}(x) \sim \frac{1}{4 \pi^2x^2} \sim \sum_{n=1}^{\infty} \frac{1}{(2 \pi)^4} \int  \frac{ m^{-2}_{Rn} R^{(2)}_{Rn}  \rho^{-1}_{1}(m^2_{Rn})}{p^2+m^{2}_{Rn }  }  e^{ip \cdot x} d^4p
\eea
enters the three-point vertex, with
\bea \label{R}
\sum_{n=1}^{\infty}    R^{(2)}_{Rn}  m^{-2}_{Rn} \rho^{-1}_1(m_{Rn}^2) \sim 1
\eea
Thus  for conserved currents of twist $2$ and dimension $3$ the existence of the conformal limit in the $UV$ sets only a weak constraint on the residues of the poles in the Kallen-Lehmann representation of the coefficient function in the $OPE$. \par 


\section{Acknowledgements}

We gave a talk about this paper at the conference: \emph{HP2 : High Precision for Hard Processes}, September 3-5, (2014), associated to the workshop:  \emph{Prospects and Precision at the Large Hadron Collider at 14 TeV}, at the Galileo Galilei Institute for Theoretical Physics, Florence, Italy, where this paper was completed. \par
We would like to thank the organizers, Daniel de Florian, Sven Moch,
Guido Montagna, Fulvio Piccinini, Dimitri Colferai for the kind invitation, the participants to the conference and the workshop for the friendly atmosphere and the scientific discussions, and the Galileo Galilei Institute and INFN for financial support. \par 
We would like to thank Riccardo Barbieri, Giacomo De Palma, Dario Francia,  Augusto Sagnotti, Massimo Taronna, at Scuola Normale Superiore, Pisa, where most of this paper was worked out, for clarifying  discussions about large-$N$ $QCD$ during the recent years.

\end{document}